\documentclass[usegraphicx,usenatbib]{mn2e}

\usepackage[total={17.8cm,24.0cm},centering]{geometry}
\usepackage{times}

\newcommand{\ud}{\mathrm{d}}
\newcommand{\apj}{ApJ}           
\newcommand{\sauron}{\texttt{SAURON}}
\newcommand{\kms}{\hbox{km s$^{-1}$}}
\newcommand{\re}{\hbox{$R_{\rm e}$}}
\newcommand{\se}{\hbox{$\sigma_{\rm e}$}}
\newcommand{\vs}{\hbox{$V/\sigma$}}
\newcommand{\vse}{\hbox{$(V/\sigma,\varepsilon)$}}
\newcommand{\plotone}[1]{\includegraphics[width=\columnwidth]{#1}}
\newcommand{\refsec}[1]{Section~\ref{#1}}
\newcommand{\reffig}[1]{Fig.~\ref{#1}}
\newcommand{\refeq}[1]{equation~(\ref{#1})}

\title[The SAURON project -- X]
{The SAURON project -- X. The orbital anisotropy of elliptical and lenticular galaxies: revisiting the \vse\ diagram with integral-field stellar kinematics}

\author[Cappellari et al.]{Michele Cappellari,$^{1,3}$\thanks{E-mail:
mxc@astro.ox.ac.uk}
Eric Emsellem,$^2$
R.\ Bacon,$^2$
M.\ Bureau,$^3$
Roger L.\ Davies,$^3$
\newauthor
P.~T.\ de Zeeuw,$^1$
Jes\'us Falc\'on-Barroso,$^{1,4}$
Davor Krajnovi\'c,$^3$
Harald Kuntschner,$^5$
\newauthor
Richard M.\ McDermid,$^1$
Reynier F.\ Peletier,$^6$
Marc Sarzi,$^7$
\newauthor
Remco C.~E.\ van den Bosch,$^1$
and Glenn van de Ven$^{1,8,9}$\\
$^1$Leiden Observatory, Leiden University, Postbus 9513, 2300 RA Leiden, The Netherlands\\
$^2$Centre de Recherche Astrophysique de Lyon, 9~Avenue Charles
    Andr\'e, 69230 Saint Genis Laval, France\\
$^3$ Sub-Department of Astrophysics, University of Oxford, Denys Wilkinson Building, Keble Road, Oxford OX1 3RH\\
$^4$European Space and Technology Centre, Postbus 299, 2200 AG Noordwijk, The Netherlands\\
$^5$Space Telescope European Coordinating Facility, European Southern
    Observatory, Karl-Schwarzschild-Str~2, 85748 Garching, Germany\\
$^6$Kapteyn Astronomical Institute, Postbus 800, 9700 AV Groningen,
    The Netherlands\\
$^7$Centre for Astrophysics Research, University of Hertfordshire, Hatfield, Herts AL 10 9AB\\
$^8$Department of Astrophysical Sciences, Peyton Hall, Princeton, NJ 08544, USA\\
$^9$Institute for Advanced Study, Einstein Drive, Princeton, NJ 08540, USA}

\pagerange{\pageref{firstpage}--\pageref{lastpage}} \pubyear{2007}

\begin{document}
\label{firstpage}
\maketitle
%
%
\begin{abstract}
We analyse the orbital distribution of elliptical (E) and lenticular (S0)  galaxies using \sauron\ integral-field stellar kinematics within about one effective (half light) radius. We construct the anisotropy diagram, which relates the ratio of the ordered and random motion in a galaxy (\vs), to its observed ellipticity ($\varepsilon$), for the 48 E/S0 galaxies from the \sauron\ survey. For a subsample of 24 galaxies consistent with axisymmetry, we use three-integral axisymmetric Schwarzschild dynamical models to recover the detailed orbital distribution and we find good agreement with the anisotropy derived from the \vse\ diagram. In the companion Paper~IX we show that the early-type galaxies can be subdivided into two classes of systems with or without a significant amount of specific stellar angular momentum. Here we show that the two classes have different distributions on the \vse\ diagram. The slow rotators are more common among the most massive systems and are generally classified as E from photometry alone. Those in our sample tend to be fairly round ($\varepsilon\la0.3$), but can have significant kinematical misalignments, indicating that as a class they are moderately triaxial, and span a range of anisotropies ($\delta\la0.3$). The fast rotators are generally fainter and are classified either E or S0. They can appear quite flattened ($\varepsilon\la0.7$), do not show significant kinematical misalignments (unless barred or interacting), indicating they are nearly axisymmetric, and span an even larger range of anisotropies ($\delta\la0.5$). These results are confirmed when we extend our analysis to 18 additional E/S0 galaxies observed with \sauron. The dynamical models indicate that the anisotropy inferred from the \vse\ diagram is due to a flattening of the velocity ellipsoid in the meridional plane ($\sigma_R>\sigma_z$), which we quantify with the $\beta$ anisotropy parameter. We find a trend of increasing $\beta$ for intrinsically flatter galaxies. A number of the fast rotators show evidence for containing a flattened, kinematically distinct component, which in some cases counter rotates relative to the main galaxy body. These components are generally more metal rich than the galaxy body. All these results support the idea that fast rotators are nearly oblate and contain disk-like components. The role of gas must have been important for their formation. The slow rotators are weakly triaxial. Current collisionless merger models seem unable to explain their detailed observed properties.
\end{abstract}
\begin{keywords}
galaxies: elliptical and lenticular, cD --
galaxies: evolution -- galaxies: formation -- galaxies: kinematics and
dynamics -- galaxies: structure
\end{keywords}
%
%
\section{Introduction}
\label{sec:intro}

Early-type galaxies are traditionally classified into elliptical (E) and lenticular (S0) galaxies. More than thirty years ago, before observations of the stellar kinematics were available, E galaxies were thought to constitute a homogeneous class of simple oblate systems, with an isotropic velocity distribution, and in which the flattening provided a measure of the galaxy angular momentum \citep[e.~g.][]{got75}. Lenticular galaxies were considered similar to spiral galaxies, containing an isotropic bulge and a disk, but without significant evidence for gas and dust.

The revolution came in the late 70's, when it became possible to extract the stellar kinematics of bright galaxies. To characterise the degree of ordered rotation in galaxies the anisotropy diagram was introduced, which relates the ratio between the ordered and random motion (\vs) in a galaxy to its observed flattening ($\varepsilon$). The small observed values of \vs, for a sample of 13 bright E galaxies, was interpreted as evidence that these systems are inconsistent with being oblate systems, close to isotropic and supported by rotation \citep{ber75,ill77,bin78}. As rotation was not needed to flatten the systems, the observations could also be explained by assuming that E galaxies were triaxial and supported by orbital anisotropy \citep{bin76,bin78}. The observations were later extended to a sample of 11 fainter E galaxies \citep{dav83} and bulges of 9 barred \citep{kor82b} and 8 unbarred \citep{ki82} spiral galaxies. From the resulting compilation of a sample of 50 Es and bulges, over two orders of magnitude in luminosity, it was found that, contrary to the brighter systems, the fainter galaxies and the spiral bulges were rotating fast and their location on the \vse\ diagram was consistent with oblate isotropic rotators \citep{dav83}.

It was later discovered that the fast rotating galaxies also tend to have disky isophotes, while the slow rotating ones have boxy isophotes \citep{ben88,ben89}. One possibility to explain this connection between photometry and kinematics was to assume the fast rotating galaxies all contained disks seen at various inclinations on top of a spheroidal stellar component \citep[e.~g.][]{rix90,rix99}. Observations at different scale heights on edge-on disk galaxies however showed that bulges themselves are rotating fast \citep{ki82}. The slow rotation of the brighter Es could be equally due to either radial or tangential anisotropy. To address this point \citet{van91} used two-integral Jeans dynamical modelling to analyse a compilation of long slit kinematics of 37 bright Es. He concluded that Es as a class are {\em radially} anisotropic. This appeared consistent with what had been observed in the outer parts of $N$-body simulations of collisionless collapse \citep{van82}. The difference between the different types of spheroidal systems could be explained in the context of galaxy formation as due to the relative importance of gaseous dissipation \citep{ben92}.

The whole picture was summarised by \citet{kor96} who proposed a revision of the standard photometric classification \citep{hub36} of early-type galaxies into E and S0. They suggested that there was a dichotomy between two classes of early-type galaxies: (i) on one side the bright, boxy, slow rotating, and radially anisotropic E galaxies, with a break between a steep outer surface brightness profile and a more shallow nuclear cuspy core, and (ii) on the other side the generally fainter, disky, fast rotating and isotropic disky-E and S0, without clear breaks between the outer and nuclear profiles.

More recently a systematic investigation of the anisotropy of a sample of 21 luminous, nearly round and slowly rotating elliptical galaxies was performed using spherical models by \citet{kro00} and \citet{ger01}. They found that ellipticals are only moderately radially anisotropic. \citet{geb03} constructed more general axisymmetric models and explored the anisotropy of a sample of 12 early-type galaxies, which included flattened objects. They found a range of anisotropy and no obvious trend (except near the nuclear supermassive black holes), but they also found the roundest galaxies to be generally only weakly anisotropic. However the models in these studies were generally fitted to the kinematics extracted along a few long-slit positions.

The introduction of panoramic integral-field spectrographs on large telescopes, combined with the advances in the dynamical modelling techniques, has opened the possibility for a reanalysis of the orbital distribution in early-type galaxies. The goal of this paper is to use \sauron\ \citep[][hereafter Paper~I]{bac01} integral-field stellar kinematics to derive the \vse\ anisotropy diagram for the \sauron\ sample \citep[][hereafter Paper~II]{dez02} in a new way that can be rigorously interpreted with the formalism of \citet{bin05}. The results are analysed making use of a new classification scheme that we introduce in a companion paper \citep[hereafter Paper~IX]{ems06}. The anisotropy derived from the \vse\ diagram is interpreted and tested using general axisymmetric dynamical models for a subsample of the survey galaxies.

The galaxy sample is described in Section~2. In Section~3 we describe the dynamical modelling technique and in Section~4 we present our modelling results. In Section~5 we present the \vse\  diagram obtained from the \sauron\ data, while in Section~6 we compare our results with previous works. Finally in Section~7 we draw some conclusions.

\section{Sample and Data}
\label{sec:sample}

Two different subsamples are used in this paper. (i) The main galaxy sample (Table~\ref{tab2}) is composed of the 48 E and S0 galaxies of the \sauron\ survey (Paper~II). The \vse\ diagram of this sample is presented in \refsec{sec:vs_lines}. (ii) For half of the survey galaxies we constructed axisymmetric dynamical models to interpret the \sauron\ \vse\ diagram. This modelling subsample of 24 galaxies (Table~\ref{tab1}) is the same used in \citet[][hereafter Paper~IV]{cap06}.

\begin{table*}
\caption{Measured parameters for the 48 E/S0 of the \sauron\ sample.}
\tabcolsep=3pt
\begin{tabular}{llcccccccccccc}
\hline
Galaxy Name & Type & T & \re\ & $R_{\rm max}/\re$ & $\varepsilon$ & $\sqrt{\langle V^2 \rangle}$ & $\sqrt{\langle\sigma^2 \rangle}$ & $(\vs)_{\rm e}$ & PA$_{\rm phot}$ & PA$_{\rm kin}$ & $\Delta$PA$_{\rm kin}$ & $\lambda_{\re}$ & Rotator   \\
            &      &   & (arcsec)   &                         &                     &  \kms                        & \kms                             &                 & (deg)      & (deg) & (deg)  &   & \\
      (1)  & (2)           & (3)   & (4)     & (5)    & (6)   & (7)   & (8)    & (9)    & (10)   & (11) & (12) & (13) & (14) \\
\hline
NGC~474    & S0$^0$(s)     & -2.0 &  29   &  0.71   &  0.11           &  31  & 146  & 0.21 &  14.0 & 152.0 & 4.5 & 0.20 & F \\
NGC~524    & S0$^+$(rs)    & -1.2 &  51   &  0.61   &  0.05           &  64  & 222  & 0.29 &  48.4 &  40.0 & 1.0 & 0.28 & F \\
NGC~821    & E6?           & -4.8 &  39   &  0.62   &  0.40           &  48  & 182  & 0.26 &  32.2 &  31.0 & 1.5 & 0.26 & F \\
NGC~1023   & SB0$^-$(rs)   & -2.7 &  48   &  0.56   &  0.33           &  62  & 176  & 0.35 &  87.0 &  89.0 & 1.0 & 0.38 & F \\
NGC~2549   & S0$^0$(r)sp   & -2.0 &  20   &  1.16   &  0.49           &  66  & 119  & 0.56 &   1.2 &   1.0 & 0.5 & 0.54 & F \\
NGC~2685   & (R)SB0$^+$pec & -1.1 &  20   &  1.34   &  0.62           &  62  &  71  & 0.88 &  36.6 &  37.0 & 1.5 & 0.72 & F \\
NGC~2695   & SAB0$^0$(s)   & -2.1 &  21   &  0.96   &  0.29           &  87  & 162  & 0.54 & 169.3 & 175.0 & 1.0 & 0.56 & F \\
NGC~2699   & E:            & -5.0 &  14   &  1.41   &  0.15           &  48  & 112  & 0.43 &  50.2 &  50.0 & 3.0 & 0.45 & F \\
NGC~2768   & E6:           & -4.3 &  71   &  0.39   &  0.38           &  51  & 209  & 0.24 &  94.7 &  95.0 & 1.5 & 0.27 & F \\
NGC~2974   & E4            & -4.7 &  24   &  1.04   &  0.37           & 127  & 180  & 0.70 &  43.5 &  43.0 & 0.5 & 0.60 & F \\
NGC~3032   & SAB0$^0$(r)   & -1.8 &  17   &  1.19   &  0.11           &  23  &  87  & 0.27 &  94.6 &  91.0 & 10. & 0.42 & F \\
NGC~3156   & S0:           & -2.4 &  25   &  0.80   &  0.47           &  42  &  48  & 0.88 &  49.4 &  50.0 & 3.5 & 0.71 & F \\
NGC~3377   & E5-6          & -4.8 &  38   &  0.53   &  0.46           &  57  & 117  & 0.49 &  41.3 &  46.0 & 1.0 & 0.47 & F \\
NGC~3379   & E1            & -4.8 &  42   &  0.67   &  0.08           &  28  & 198  & 0.14 &  67.9 &  72.0 & 2.0 & 0.14 & F \\
NGC~3384   & SB0$^-$(s):   & -2.7 &  27   &  0.75   &  0.20           &  57  & 131  & 0.44 &  53.0 &  48.0 & 1.5 & 0.41 & F \\
NGC~3414   & S0~pec        & -2.1 &  33   &  0.60   &  0.21           &  19  & 206  & 0.09 & 179.9 &   1.0 & 9.5 & 0.06 & S \\
NGC~3489   & SAB0$^+$(rs)  & -1.3 &  19   &  1.05   &  0.29           &  52  &  78  & 0.67 &  71.2 &  73.0 & 1.0 & 0.60 & F \\
NGC~3608   & E2            & -4.8 &  41   &  0.49   &  0.18           &   8  & 179  & 0.05 &  79.3 &  85.0 & 8.5 & 0.04 & S \\
NGC~4150   & S0$^0$(r)?    & -2.1 &  15   &  1.39   &  0.30           &  37  &  64  & 0.58 & 147.0 & 147.0 & 4.5 & 0.58 & F \\
NGC~4262   & SB0$^-$(s)    & -2.7 &  10   &  2.06   &  0.22$^\dagger$ &  40  & 166  & 0.24 & 154.0 & 150.0 & 1.5 & 0.25 & F \\
NGC~4270   & S0            & -1.9 &  18   &  1.09   &  0.50           &  46  & 114  & 0.40 & 107.5 & 102.0 & 3.0 & 0.45 & F \\
NGC~4278   & E1-2          & -4.8 &  32   &  0.82   &  0.12           &  44  & 228  & 0.19 &  16.7 &  12.0 & 0.5 & 0.15 & F \\
NGC~4374   & E1            & -4.2 &  71   &  0.43   &  0.15           &   7  & 282  & 0.03 & 128.2 & 141.0 & 1.5 & 0.02 & S \\
NGC~4382   & S0$^+$(s)pec  & -1.3 &  67   &  0.38   &  0.19           &  31  & 192  & 0.16 &  29.4 &  23.0 & 1.5 & 0.16 & F \\
NGC~4387   & E             & -4.8 &  17   &  1.16   &  0.40           &  34  &  89  & 0.39 & 141.5 & 151.0 & 6.0 & 0.41 & F \\
NGC~4458   & E0-1          & -4.8 &  27   &  0.74   &  0.12           &  10  &  84  & 0.12 &   4.5 &  24.0 & 17  & 0.05 & S \\
NGC~4459   & S0$^+$(r)     & -1.4 &  38   &  0.71   &  0.17           &  66  & 146  & 0.45 & 102.7 & 100.0 & 0.5 & 0.44 & F \\
NGC~4473   & E5            & -4.7 &  27   &  0.92   &  0.41           &  41  & 188  & 0.22 &  93.7 &  92.0 & 1.0 & 0.19 & F \\
NGC~4477   & SB0(s):?      & -1.9 &  47   &  0.43   &  0.24$^\dagger$ &  33  & 158  & 0.21 &  64.0 &  70.0 & 2.5 & 0.22 & F \\
NGC~4486   & E0-1$^+$pec   & -4.3 & 105   &  0.29   &  0.04           &   7  & 306  & 0.02 & 158.2 & 113.2 & 45. & 0.02 & S \\
NGC~4526   & SAB0$^0$(s)   & -1.9 &  40   &  0.66   &  0.37           & 103  & 189  & 0.54 & 112.8 & 111.0 & 0.5 & 0.47 & F \\
NGC~4546   & SB0$^-$(s):   & -2.7 &  22   &  0.94   &  0.45           &  97  & 161  & 0.60 &  75.0 &  79.0 & 0.5 & 0.60 & F \\
NGC~4550   & SB0$^0$:sp    & -2.0 &  14   &  1.45   &  0.61           &  12  & 116  & 0.10 & 178.3 & 178.0 & 1.0 & 0.09 & S$^\star$ \\
NGC~4552   & E0-1          & -4.6 &  32   &  0.63   &  0.04           &  13  & 257  & 0.05 & 125.3 & 113.0 & 5.0 & 0.05 & S \\
NGC~4564   & E             & -4.8 &  21   &  1.02   &  0.52           &  76  & 131  & 0.58 &  48.6 &  49.0 & 2.0 & 0.59 & F \\
NGC~4570   & S0~sp         & -2.0 &  14   &  1.43   &  0.60           &  81  & 152  & 0.53 & 159.3 & 159.0 & 0.5 & 0.56 & F \\
NGC~4621   & E5            & -4.8 &  46   &  0.56   &  0.34           &  52  & 207  & 0.25 & 163.3 & 165.0 & 0.5 & 0.27 & F \\
NGC~4660   & E             & -4.7 &  11   &  1.83   &  0.44           &  79  & 163  & 0.49 &  96.8 &  98.0 & 0.5 & 0.47 & F \\
NGC~5198   & E1-2:         & -4.7 &  25   &  0.80   &  0.12           &  12  & 185  & 0.07 &  15.3 &  58.0 & 21. & 0.06 & S \\
NGC~5308   & S0$^-$~sp     & -2.0 &  10   &  2.04   &  0.60           &  86  & 192  & 0.45 &  58.5 &  59.0 & 1.0 & 0.48 & F \\
NGC~5813   & E1-2          & -4.8 &  52   &  0.53   &  0.15           &  32  & 223  & 0.14 & 134.5 & 151.0 & 2.5 & 0.06 & S \\
NGC~5831   & E3            & -4.8 &  35   &  0.67   &  0.15           &  11  & 151  & 0.08 & 122.8 & 101.0 & 10. & 0.05 & S \\
NGC~5838   & S0$^-$        & -2.7 &  23   &  0.87   &  0.34           & 110  & 216  & 0.51 &  41.7 &  39.0 & 0.5 & 0.52 & F \\
NGC~5845   & E:            & -4.8 &  4.6  &  4.45   &  0.35           &  81  & 226  & 0.36 & 143.2 & 141.0 & 2.0 & 0.36 & F \\
NGC~5846   & E0-1          & -4.7 &  81   &  0.29   &  0.07           &   7  & 240  & 0.03 &  75.2 & 126.0 & 5.0 & 0.02 & S \\
NGC~5982   & E3            & -4.8 &  27   &  0.94   &  0.30           &  19  & 234  & 0.08 & 108.9 & 114.0 & 4.0 & 0.09 & S \\
NGC~7332   & S0~pec~sp     & -2.0 &  11   &  1.91   &  0.42           &  38  & 116  & 0.32 & 159.8 & 152.0 & 1.5 & 0.39 & F \\
NGC~7457   & S0$^-$(rs)?   & -2.6 &  65   &  0.39   &  0.44           &  38  &  62  & 0.62 & 125.5 & 124.0 & 4.0 & 0.57 & F \\
\hline
\end{tabular}
\begin{minipage}{17.8cm}
Notes:
(1) NGC number.
(2) Morphological type from \citet[hereafter RC3]{dev91}.
(3) Numerical morphological T-type (LEDA. E: $T \le -3.0$, S0:  $-3.0 < T \le -0.5$).
(4) Effective (half-light) radius \re\ measured in the $I$-band from HST/WFPC2 $+$ MDM images as described in Paper~IV. Comparison with the RC3 values, for the 46 galaxies in common, shows an rms scatter of 20\%.
(5) Ratio between the maximum radius $R_{\rm max}$ sampled by the kinematical observations and \re. We defined $R_{\rm max}\equiv\sqrt{S/\pi}$, where $S$ is the area on the sky sampled by the \sauron\ observations.
(6) Luminosity-weighted average ellipticity. This was computed from the ellipse of inertia of the galaxy surface brightness inside an isophote enclosing an area $A=\pi\re^2$, or within the largest isophote fully contained within the \sauron\ field, whichever is smaller.
(7) Luminosity-weighted squared velocity within an ellipse of area $A$, ellipticity $\varepsilon$, and PA given in column [10], or within the largest similar ellipse fully contained within the \sauron\ field, whichever is smaller.
(8) Luminosity-weighted squared velocity dispersion inside the same ellipse as in column [7].
(9) Luminosity-weighted \vs\ ratio within 1\re. This is the ratio of columns [7] and [8]. See equation~[\ref{eq:vsigma}] for a definition of these quantities.
(10) Large scale global luminosity-weighted PA of the photometric major axis.
(11) PA of the global kinematic major axis within the \sauron\ field (direction where $|V|$ is maximum, see text for details).
(12) Error in the kinematic major axis of column [11].
(13) $\lambda_R$ (see Paper~IX) measured within the same ellipse as in column [7] (about 1\re).
(14) Galaxy classification from Paper~IX: F$=$fast-rotator ($\lambda_{\re}>0.1$), S$=$slow-rotator ($\lambda_{\re}\le0.1$).

$^\star$ This galaxy is a special slow-rotators. It appears axisymmetric and disk-like as a fast-rotator, but contains two counterrotating disks (\refsec{sec:understand}).

$^\dagger$ These two galaxies show nearly face-on bars. The ellipticity is the one of the outer disk, which also defines the photometric PA.
\end{minipage}
\label{tab2}
\end{table*}

\begin{table}
\centering
\caption{Anisotropy parameters for the 24 modelled galaxies.}
\tabcolsep=5pt
\begin{tabular}{lcccccc}
\hline
Galaxy Name & $i$ & $\beta_r$ & $\beta$ & $\gamma$ & $\delta$ & $\delta_{(\vs)}$  \\
            &(deg)&         &         &          &          &                   \\
        (1) & (2) & (3)     & (4)     & (5)      &  (6)     &   (7)          \\
\hline
NGC~524   &  19  &  0.06 &  0.17 & -0.04 & 0.19 &   0.19   \\
NGC~821   &  90  &  0.16 &  0.21 &  0.04 & 0.20 &   0.30   \\
NGC~2974  &  57  & -0.20 &  0.13 & -0.30 & 0.24 &   0.20   \\
NGC~3156  &  68  &  0.17 &  0.39 &  0.19 & 0.33 &   0.19   \\
NGC~3377  &  90  &  0.07 &  0.28 &  0.08 & 0.25 &   0.23   \\
NGC~3379  &  90  &  0.11 &  0.06 &  0.06 & 0.03 &   0.04   \\
NGC~3414  &  90  & -0.12 &  0.06 & -0.12 & 0.11 &   0.17   \\
NGC~3608  &  90  &  0.04 &  0.10 & -0.06 & 0.13 &   0.15   \\
NGC~4150  &  52  & -0.01 &  0.32 & -0.12 & 0.36 &   0.22   \\
NGC~4278  &  90  & -0.02 &  0.11 & -0.17 & 0.18 &   0.06   \\
NGC~4374  &  90  &  0.11 &  0.10 &  0.05 & 0.08 &   0.12   \\
NGC~4458  &  90  & -0.26 & -0.01 & -0.23 & 0.09 &   0.08   \\
NGC~4459  &  47  &  0.10 &  0.05 &  0.11 & 0.00 &   0.02   \\
NGC~4473  &  73  & -0.21 &  0.18 & -0.50 & 0.34 &   0.37   \\
NGC~4486  &  90  &  0.24 &  0.11 &  0.22 & 0.00 &   0.03   \\
NGC~4526  &  79  &  0.11 &  0.11 &  0.09 & 0.06 &   0.09   \\
NGC~4550  &  84  & -0.37 &  0.43 & -0.87 & 0.60 &   0.56   \\
NGC~4552  &  90  & -0.06 &  0.01 & -0.03 & 0.02 &   0.03   \\
NGC~4621  &  90  & -0.04 &  0.11 & -0.17 & 0.18 &   0.24   \\
NGC~4660  &  70  &  0.02 &  0.27 & -0.11 & 0.30 &   0.30   \\
NGC~5813  &  90  &  0.17 &  0.18 &  0.21 & 0.08 &   0.10   \\
NGC~5845  &  90  &  0.24 &  0.23 &  0.18 & 0.15 &   0.19   \\
NGC~5846  &  90  &  0.17 &  0.09 &  0.17 & 0.01 &   0.06   \\
NGC~7457  &  64  &  0.03 &  0.38 &  0.04 & 0.37 &   0.31   \\
\hline
\end{tabular}
\begin{minipage}{8cm}
Notes:
(1) NGC number. (2) Inclination from paper~IV. (3) anisotropy parameter $\beta_r$ measured in spherical coordinates from the solution of the dynamical models as defined in equation~[\ref{eq:beta_r}]. (4)--(6) anisotropy parameters $\beta$, $\gamma$ and $\delta$ determined in cylindrical coordinates from the solution of the dynamical models. The parameters are defined in equations~[\ref{eq:beta},\ref{eq:gamma}] and equation~[\ref{eq:delta}] respectively. (7) Anisotropy parameter $\delta$ as measured using the \vse\ diagram.
\end{minipage}
\label{tab1}
\end{table}

All the galaxies used have \sauron\ integral-field spectroscopy out to about one effective (half light) radius (\re). The \sauron\ data were reduced as described in \citet[][hereafter Paper~III]{ems04}. However there are some differences between the kinematics presented in Paper~III and the one used for this work:
(i) To provide a tight constraint to the dynamical models we extracted the Gauss-Hermite (G-H) moments \citep{van93,ger93} of the line-of-sight stellar velocity distribution (LOSVD) up to $h_3$--$h_6$, using the penalised pixel-fitting method \citep[pPXF,][]{cem04};
(ii) To measure the mean velocity $V$ and the velocity dispersion $\sigma$ to be used in the \vse\ diagram we did not fit the higher order G-H moments. We verified using pPXF on semianalitic dynamical models that this generally provides a better approximation to the first and second velocity moment of the LOSVD, which appear in the tensor-virial equations from which the \vse\ diagram is constructed;
(iii) To reduce the influence of template mismatch we updated the library of templates which is fitted together with the kinematics by pPXF. In particular we constructed the optimal template using the 985 stars of the MILES library \citep{san06}, from which $\approx15$ stars are selected by the program to provide a detailed fit to each galaxy spectrum. The use of this new library can reduce the rms scatter in the residuals of the pPXF fit by up to a factor $\approx3$, when the spectra have negligible Poissonian noise. One can expect a reduction of the systematic errors in the G-H moments by up to the same factor (see Fig.~B3 of Paper~III). In practice our results are very similar to the ones presented in Paper~III, but for some of the most massive galaxies the $h_4$ values are here significantly lower \citep[see also][]{sha06}.

Together with the HST/WFPC2 photometry in the $I$-band, wide-field ground-based MDM photometry (Falcon-Barroso et al.\ in preparation) is also available for all the modelled galaxies. We used this set of photometric data to parameterise the stellar density distribution in our models according to the multi-Gaussian expansion (MGE) method \citep{ems94,cap02}. The parameters of the MGE models are given in Paper~IV. The MDM photometry was also used to determine the \re\ of the sample galaxies.

An additional set of 18 E/S0 galaxies has been independently observed with a similar SAURON setup in the course of various other projects. These objects will be treated as ``specials'', most having features which motivated a specific observation, and will be only mentioned in \refsec{sec:specials} to strengthen the results obtained from the main survey. As in Paper~II the galaxies are classified as either E or S0 from the LEDA morphological type ($T\le-0.5$; \citealt{pat03}). The kinematics of these galaxies were not presented in Paper~III, and will be presented elsewhere.

\section{Three-integral dynamical modelling}
\label{sec:models}

The models we study in this paper were presented in Paper~IV. The stationary and axisymmetric dynamical modelling technique that we use is based on the \citet{sch79} numerical orbit-superposition method, which was extended to fit kinematical observables \citep{ric88,rix97,van98}. The implementation of the method that we adopt in this paper was optimised for use with integral-field data and is described in Paper~IV. Similar axisymmetric implementations were developed by other groups \citep{geb03,val04,tho04}.

For a given stationary gravitational potential, the stellar dynamics of a galaxy is uniquely defined by the orbital distribution function (DF), which describes the velocities of the stars at every position in the galaxy. As the stellar orbits in a stationary potential conserve at most three isolating integrals of motion, the DF can be written as a function of the three integrals, or any other equivalent parameterization of them. From dimensionality arguments this implies that the three-dimensional DF cannot be recovered without at least the knowledge of the LOSVD at every spatial position $(x',y')$ on the galaxy image on the sky, which also constitutes a three-dimensional quantity. For an axisymmetric edge-on galaxy, with a {\em given} potential, this knowledge of the LOSVD seems likely sufficient for a unique recovery of the DF, which may not be positive everywhere if the assumed potential is wrong \citep[see Section~3 of][]{val04}. An example of the need for two-dimensional kinematics to constrain the DF is given in \citet{cmd05}. In \citet{kra05} and \citet{van07} we verified the ability of our implementation of the Schwarzschild method to recover the DF and the internal velocity moments for both two and three-integral realistic galaxy model, even from incomplete radial coverage.

An implicit assumption of the dynamical models is that the luminosity density, as can be obtained by deprojecting the galaxy surface brightness, provides a good description of the shape of the total density. This is also a key assumption of the \vse\ diagram. It implies that either dark matter provides a small contribution to the total matter in the regions we study (inside 1\re), or that its shape is similar to that of the luminous matter. Evidence from  dynamical modelling \citep[e.g.][]{ger01,cap06} and gravitational lensing \citep[e.g.][]{tre04,rus05,koo06} suggest that both assumptions are reasonably well justified in real galaxies.

In \citet{kra05} and in Paper~IV we showed that there is evidence for the inclination to be possibly degenerate, even with the knowledge of the LOSVD at all spatial positions. As discussed in Paper~IV, we adopted for our models the assumption-dependent inclination derived by fitting two-integral Jeans models. This inclination appears to provide values in agreement with the geometry of dust/gas disks when they are present.
However for 14 of the 24 modelled galaxies the inclination is already constrained by arguments independent from the dynamics. In fact nine galaxies (NGC~821, NGC~3156, NGC~3377, NGC~4473, NGC~4550, NGC~5845, NGC~4621, NGC~4660, NGC~5845), show significantly disky isophotes and require Gaussians flatter than $q'\la0.4$ in their MGE models (see Paper~IV). Under the assumption of axisymmetry, this implies they cannot be too far from edge-on \citep[$i\ga70$; see][\S~2.2.2]{cap02}. The inclination of five of the remaining galaxies can be derived from the geometry of a gas or dust disk, assuming it is in equilibrium in the equatorial plane of an oblate galaxy (NGC~524, NGC~2974, NGC~4150, NGC~4459, NGC~4526), and it agrees with the dynamically determined inclination (Paper~IV).

\section{Modelling results}
\label{sec:modeling}

\subsection{Internal velocity moments}
\label{sec:momemnts}

\begin{figure*}
    \includegraphics[width=0.75\textwidth]{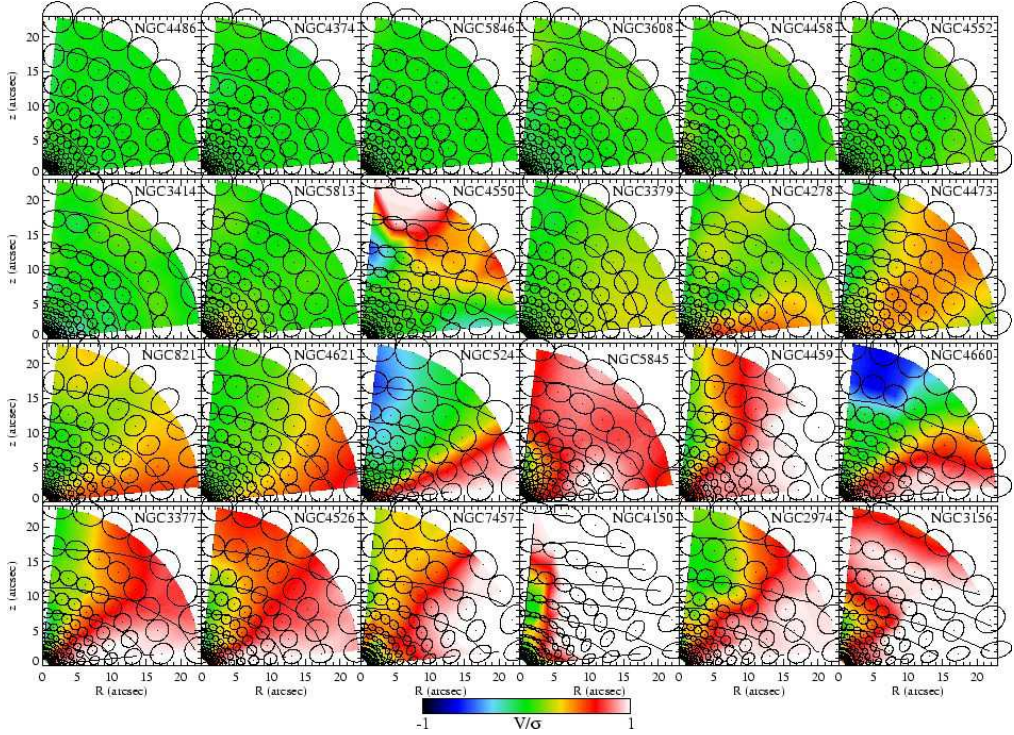}
    \caption{Internal moments of the modelling subsample. The colours visualise the ratio $\langle V \rangle/\bar{\sigma}$ of the first and second internal moment of the velocity, in the positive quadrant of the meridional plane $(R,z)$, for the 24 galaxies of the modelling subsample. The contours of the intrinsic density are overlaid. The ellipses are cross sections of the velocity ellipsoid with the $(v_R,v_z)$ plane at different locations (one of the principal axes of the ellipsoid is orthogonal to the plane). The objects are sorted, from left to right and from top to bottom, according to the parameter $\lambda_{\re}$ (Table~\ref{tab2}) which is related to the specific angular momentum per unit mass. The first nine galaxies are classified as slow-rotators  ($\lambda_{\re}<0.1$) and the remaining fifteen as fast-rotators ($\lambda_{\re}>0.1$). All panels have the same size, which roughly corresponds to the area fully covered by the observed \sauron\ kinematics.}
    \label{fig:rotation_all_sample}
\end{figure*}

Here we present the orbital distribution of the 24 early-type galaxies of our modelling subsample (\refsec{sec:sample}), derived from our Schwarzschild models, at the inclination determined in Paper~IV. The models provide a full description of the DF in terms of a three-dimensional set of weights, which could in principle be transformed into a distribution of stellar mass as a function of the three integrals of motion ($E$, $L_z$, and a nonclassical third integral $I_3$).
In this paper we will mainly focus on a more concise description of the main characteristics of the internal kinematics which can be obtained by computing the first and second moments of the velocity distribution. Given the assumed axial symmetry, it is sufficient to study the moments in the galaxy meridional plane $(R,z)$.

The results of this section are based on a sample of galaxies which was selected to be consistent with the axisymmetric assumption. The sample contains both fast-rotating and nearly non-rotating galaxies, but it does not include strongly triaxial ones. This has to be considered when drawing statistical conclusions from the models. The results of this section are important to interpret and test the more general results we will obtain using the \vse\ diagram in \refsec{sec:vs_sauron}, for the whole \sauron\ sample which also includes significantly triaxial galaxies.

A large number of studies, in the past twenty years, has shown that elliptical galaxies display a dichotomy between the most massive galaxies, which tend to rotate slowly, are metal-rich, have a flat central luminosity profile and show evidence of triaxiality, and the less massive objects, which rotate faster, are metal-poor, have a cuspy luminosity profile, and may all be axisymmetric and contain disks \citep{dav83,ben88,kor96,fab97,lau05}.

In the companion Paper~IX we show that a clean separation of the two classes of early-type galaxies is obtained by introducing a new {\em quantitative} classification parameter $\lambda_R$ which is related to the specific angular momentum of galaxies \citep[e.g.][]{ber78,ben90} and which can be measured from integral-field stellar kinematics as
\begin{equation}
\lambda_R\equiv\frac{\langle R |V| \rangle}{\langle R \sqrt{V^2+\sigma^2} \rangle} = \frac{\sum_{n=1}^{N} F_n\, R_n |V_n|}{\sum_{n=1}^{N} F_n\,R_n  \sqrt{V_n^2+\sigma_n^2}},
\label{eq:lambda}
\end{equation}
where $F_n$ is the flux contained inside the $n$-th Voronoi bin and $V_n$ and $\sigma_n$ the corresponding measured mean velocity and velocity dispersion (see Paper IX). The numerical values of $\lambda_R$ for our sample, measured with 1\re, are given in Table~\ref{tab2}.
In this work we adopt that classification to define the galaxies in our sample with significant angular momentum per unit mass $\lambda_R>0.1$, which we call ``fast-rotators'' and the ones with negligible amount of specific angular momentum $\lambda_R<0.1$, which we define as ``slow-rotators'' (Table~\ref{tab2}). Our kinematic classification has a similar spirit to the one proposed by \citet{kor96}, but contrary to the latter, which was based on the galaxy isophotal shape, our classification is quite robust against projection effects.

In \reffig{fig:rotation_all_sample} we plot the internal moments in the meridional plane of our subsample of 24 modelled galaxies. The ellipses represent the cross section of the velocity ellipsoids at various positions in the galaxy body, while the colours visualise the ratio $\langle v_\phi \rangle/\bar{\sigma}$ between the mean velocity around the symmetry axis and the local mean velocity dispersion $\bar{\sigma}^2=(\sigma_r^2+\sigma_\theta^2+\sigma_\phi^2)/3$, where $\sigma_\phi^2=\langle v_\phi^2 \rangle - \langle v_\phi \rangle^2$. Here $(r,\theta,\phi)$ are the standard spherical coordinates ($\phi$ is the azimuthal angle around the galaxy symmetry axis). The quantity $\langle v_\phi \rangle/\bar{\sigma}$ is a local analogue, inside the galaxy body, of the global quantity $V/\sigma$ which measures the importance of rotation with respect to the random motions. In the plot the galaxies were sorted according to the $\lambda_R$ parameter. The first nine galaxies are classified as slow-rotators, while the remaining fifteen are fast-rotators.

The slow rotators in this modelling subsample are all intrinsically quite round, with an axial ratio of the density $q\ga0.8$. This result depends on the correctness of the assumed inclination, which in all cases is best fitted as edge-on ($i=90^\circ$). This inclination cannot be correct in all cases, however these objects are all very round in projection, which statistically implies they are intrinsically not far from spherical. The flattest slow-rotator of this subsample has an apparent axial ratio of $q'\approx0.8$ and assuming it has the mean inclination of a randomly oriented sample ($i=60^\circ$), it would still have an intrinsic axial ratio $q\approx0.7$. As we verified numerically with our models, for nearly round models the results for the orbital distribution change little with inclination, as can be understood from the obvious fact that a spherical model appears identical from any inclination. In particular we constructed models for NGC~3379, NGC~4486 and NGC~4552 at an inclination of $i=45^\circ$ and found virtually unchanged anisotropy results. The slow rotators of this subsample are generally characterized by a nearly round velocity ellipsoid in the meridional plane and do not show any significant rotation. A clear exception is NGC~4550, which shows internal rotation and a strongly flattened velocity ellipsoid.

Our anisotropy results could be significantly in error if the slow rotators of our modelling subsample were intrinsically flattened and nearly face-on objects ($i\sim0$). However the small apparent ellipticity of the slow-rotators is also generally true for the whole \sauron\ sample. That sample includes some flatter and clearly triaxial slow-rotators, however the smallest apparent axial ratio of any slow-rotator is still $q'\approx0.7$.  Statistically this implies the ratio of the shortest and longest axis of the intrinsic density is $c/a\ga0.7$. In Paper~IX we show that the slow rotators are truly different from the fast-rotators and do not simply {\em appear} different due to projection. All this makes it very unlikely that any of the slow-rotators we modelled is a flat system seen nearly face-on and implies our anisotropy results for the subsample are reliable.

The fast rotators appear to span a large range of intrinsic flattening with $q\approx0.3-0.9$. For the flat objects, the maximum contribution to rotation generally occurs on the galaxy equatorial plane, as expected, but an important exception is NGC~4473, which has a minimum of the velocity contribution on the major axis (as NGC~4550). The orbital structure of the two special cases NGC~4473 and NGC~4550 will be explained in \refsec{sec:understand}. The fast-rotators are generally significantly anisotropic, with large variations in the ratio $\sigma_r/\sigma_\theta$. As a first approximation the velocity ellipsoid tends to be aligned in polar coordinates and appears to be flattened in the same direction as the density distribution, in the sense that $\sigma_r<\sigma_\theta$ along the galaxy rotation axis, while $\sigma_r>\sigma_\theta$ on the equatorial plane. Using cylindrical coordinates in the meridional plane $(R,z)$, one may broadly describe the observed shape of the velocity ellipsoid as being generally flattened along the $z$ direction. There seems to be a tendency for the anisotropy to be stronger near the equatorial plane, especially in disk-dominated galaxies like NGC~3156, NGC~4150 and NGC~7457.

\subsection{Global velocity dispersion tensor}
\label{sec:global}

In the previous section we described the variations of the velocity ellipsoid as a function of position inside the galaxies. In this section we consider global integrated quantities for each galaxy. A classic way to quantify the global anisotropy in galaxies is by using the anisotropy parameter \citep[][\S~4.3]{bin87}
\begin{equation}
    \delta\equiv 1 - \frac{\Pi_{zz}}{\Pi_{xx}},
    \label{eq:delta}
\end{equation}
where $z$ coincides with the symmetry axis of an axisymmetric galaxy, $x$ is any fixed direction orthogonal to it and
\begin{equation}
    \Pi_{k k}=\int\rho\,\sigma_k^2\, d^3\mathbf{x},=\sum_{n=1}^{N} M_n\, \sigma_{k,n}^2,
    \label{eq:pi}
\end{equation}
with $\sigma_k$ the velocity dispersion along the direction $k$ at a given location inside the galaxy. The summation defines how we computed this quantity from our Schwarzschild models. $M_n$ is the mass contained in each of the $N$ polar bins in the meridional plane of the model, and $\sigma_{k,n}$ is the corresponding mean velocity dispersion along the direction $k$. At any location in the galaxy, the velocity ellipsoid is defined by having the principal axes along the directions which diagonalize the tensor $\mathbf{\sigma}$.
We define two additional anisotropy parameters:
\begin{equation}
    \beta\equiv 1 - \frac{\Pi_{zz}}{\Pi_{RR}},
    \label{eq:beta}
\end{equation}
describes the global shape of the velocity dispersion tensor in the $(v_R,v_z)$ plane. $\beta=0$ if the intersection of the velocity ellipsoid in the $(v_R,v_z)$ plane (as shown in \reffig{fig:rotation_all_sample}) is everywhere a circle. This is the case e.~g.\ if the DF depends only on the two classical integrals of motion $f=f(E,L_z)$, where $E$ is the energy and $L_z$ is the angular momentum parallel to the $z$-axis. If the shape of the velocity ellipsoid is constant inside the galaxy body then $\beta=1-(\sigma_z/\sigma_R)^2$. The second parameter
\begin{equation}
    \gamma\equiv 1 - \frac{\Pi_{\phi\phi}}{\Pi_{RR}}
    \label{eq:gamma}
\end{equation}
describes the global shape of the velocity dispersion tensor in a plane orthogonal to $v_z$. $\gamma=0$ when the intersection of the velocity ellipsoid with a plane orthogonal to $v_z$ axis is a circle everywhere. $\beta=\gamma=\delta=0$ for an isotropic system (spherical velocity ellipsoid everywhere).
Integrating over the azimuthal angle one finds
\begin{equation}
    \Pi_{xx}=\frac{\Pi_{RR} + \Pi_{\phi\phi}}{2},
\end{equation}
so that the three anisotropy parameters are related by
\begin{equation}
    \delta=\frac{2\beta-\gamma}{2-\gamma}.
\end{equation}
In the case $\gamma=0$ the anisotropy can be measured directly in the meridional plane and the simple relation $\beta\sim\delta$ applies.

\begin{figure}
    \centering\includegraphics[width=0.75\columnwidth]{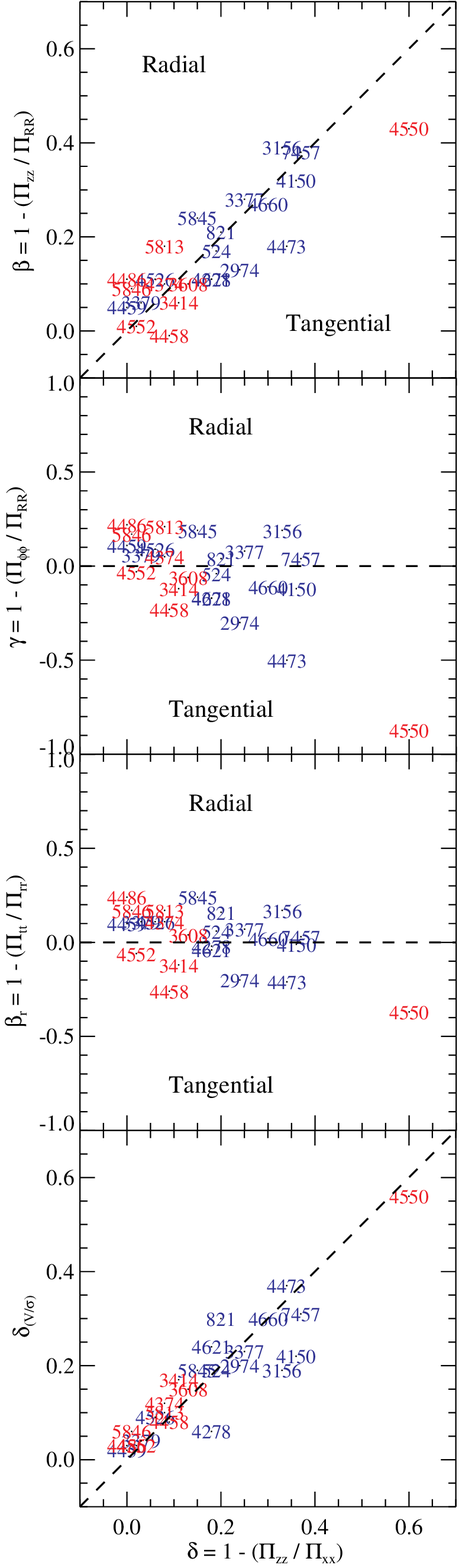}
    \caption{Global anisotropy from the dynamical models. Comparison between the anisotropy $\delta$ determined from the 24 Schwarzschild models of \reffig{fig:rotation_all_sample} and the anisotropy parameter $\beta$ ({\em Top Panel}), $\gamma$ ({\em Second Panel}) and $\beta_r$ ({\em Third Panel}). {\em Bottom Panel:} Comparison between $\delta$ from the models and $\delta_{(\vs)}$ from the \vse\ diagram of \reffig{fig:v_over_sigma}, after correction for the inclination. The red and blue labels refer to the slow and fast rotators respectively.}
    \label{fig:anisotropy}
\end{figure}

We evaluated these anisotropy parameters from the solution of the Schwarzschild models, restricting the volume integral of \refeq{eq:pi} {\em only} within the radius ($R<25''$) fully constrained by the kinematics which is shown in \reffig{fig:rotation_all_sample}. The results are shown in the top two panels of \reffig{fig:anisotropy}. In general we find that $\delta\sim\beta$ (similar to the non-classic two-integral models of \citealt{deh93}), while there is no obvious trend of $\gamma$ with increasing anisotropy $\delta$. This means that the observed anisotropy is mainly due to a flattening of the velocity dispersion tensor in the $z$ direction. However most of galaxies have $\gamma\ga0$, indicating mild radial anisotropy. Just a few galaxies have instead tangential anisotropy ($\gamma\la0$), and in particular NGC~4473 and NGC~4550 stand out (see \refsec{sec:understand}).

The near isotropy of the roundest galaxies is not surprising in this plot, in fact the anisotropy parameters are computed in cylindrical coordinates $(R,z)$, as best suited for oblate bodies. By construction all the anisotropy parameters are zero by symmetry in the spherical limit. This is a known feature of the \vse\ diagram as well, where spherical objects necessarily have zero anisotropy $\delta$. To measure the anisotropy of nearly spherical galaxies one can define a complementary anisotropy parameter, in spherical coordinates:
\begin{equation}
    \beta_r\equiv 1 - \frac{\Pi_{tt}}{\Pi_{rr}},
    \label{eq:beta_r}
\end{equation}
with
\begin{equation}
    \Pi_{tt}=\frac{\Pi_{\theta\theta} + \Pi_{\phi\phi}}{2},
\end{equation}
and $(r,\theta,\phi)$ the standard spherical coordinates. In the spherical limit, assuming the galaxy is non rotating, $\Pi_{\theta\theta} = \Pi_{\phi\phi} = \Pi_{rr}$ by symmetry.
The parameter is $\beta_r=0$ for an isotropic galaxy and is positive (negative) when the luminosity-weighted average dispersion along the radial direction is larger (smaller) than the average dispersion along any direction orthogonal to it. The plot of $\beta_r$ versus $\delta$ is shown in the third panel of \reffig{fig:anisotropy}. It confirms that the small anisotropy of the slow-rotators in the top two panels is not an artifact of the adopted coordinate system. This can also be seen qualitatively in \reffig{fig:rotation_all_sample}, which shows that the intersection of the velocity ellipsoids of the slow-rotators with the $(v_R,v_z)$ plane is nearly a circle ($\sigma_\theta\approx\sigma_r$). None of the roundish slow-rotators appear to be dominated by radial anisotropy in the regions that we observe and the inferred values of $|\beta_r|\la0.2$ imply that the average shape of the velocity ellipsoid for the roundish galaxies does not differ by more than $\approx10\%$ from a spherical shape.

The results in this section quantify what we discussed qualitatively in \refsec{sec:momemnts}, namely the fact that the slow-rotators in the modelling subsample of 24 galaxies tend to be nearly isotropic, while the fast-rotators span a large range of anisotropies. As a consistency test, in Appendix~\ref{sec:jeans} we use two-integral Jeans models to show that the slow-rotators are well reproduced by isotropic models in their central regions, while the flat fast-rotators are not.
The same considerations on the bias of the selection of this modelling subsample against strongly triaxial galaxies, that we discussed in \refsec{sec:momemnts}, also apply to the results of this section.

The anisotropy we measured was derived by fitting kinematical data which have limited spatial coverage, so the values are not necessarily representative of the galaxy as a whole. In \citet{kra05} we showed that the orbital distribution inferred from our orbital-superposition models, in the case of integral-field data with limited spatial coverage, is accurate in the region of the galaxy for which we have kinematical data, which in our case generally corresponds to 1\re. For most of the slow-rotators however, we generally do not reach 1\re\ with our \sauron\ data. This has to be kept in mind when comparing our results to the results obtained e.g.\ from numerical simulations.

\subsection{Comparison with previous modelling results}

The modelling results discussed in this paper are the first based on integral-field stellar kinematics, which is critical for the recovery of the orbital distribution of flattened galaxies. No previous results are available using a comparable technique. In the spherical limit however observations along a single slit position are in principle sufficient to uniquely recover the DF \citep{dej92}. The anisotropy we derive for the roundest galaxies can be compared with previous results obtained using spherical models.

The largest set of spherical models of galaxies in common with our sample is the one presented in \citet{kro00} and analysed in \citet{ger01}. The five galaxies in common with our modelling subsample of 24 galaxies are NGC~3379, NGC~4278, NGC~4374, NGC~4486 and NGC~5846. The galaxy with the largest apparent ellipticity is NGC~4374, which has $\varepsilon=0.15$, so our models, which take the small nonzero ellipticity into account, do not differ too much from spherical models, and we can compare the results obtained with different kinematical data and different methods.  For four of the galaxies we find that the small radial anisotropy $\beta_r\approx0.1-0.2$ that we measure does not differ by more than $\Delta\beta_r\la0.1$ from the $\beta_{\rm mean}$ (kindly provided by O. Gerhard) of \citet{ger01}, which is the unweighted mean of their radial anisotropy within 1\re (their Fig.~5). This error corresponds to a difference of just 5\% on the mean flattening of the velocity ellipsoid in these galaxies. This value likely represents the realistic systematic error one can expect from this type of measurements with current data and techniques. For one galaxy, NGC~4278, they measure a strong anisotropy $\beta_{\rm mean}\approx0.3$, while we derive near isotropy $\beta_r\approx0$. The difference could be explained by the fact that the data of \citet{kro00} for this galaxy are not very extend in radius $R\la\re/3$ and possess the lowest quality of the considered test sample.

Another systematic study of the anisotropy of a sample of 12 early-type galaxies, using an axisymmetric Schwarzschild method similar to the one adopted for the present paper was performed by \citet{geb03}. We can compare the results for the five galaxies in common that we both modelled as edge-on NGC~821, NGC~3377, NGC~3379, NGC~3608 and NGC~5845. For this we computed from our models an approximation to their mean ratio $(\sigma_r/\sigma_t)_{{\re}/4}$ (their Table~1 and 3). Contrary to our anisotropy parameter $\beta_r$, their ratio is {\em not} directly related to the shape of the velocity ellipsoid, as it includes both random and ordered azimuthal streaming (see their Section~4.7). In particular according to their definition, unless the galaxy is spherical, an isotropic model will have $(\sigma_r/\sigma_t)<1$, while a ratio of unity indicates radial anisotropy ($\beta_r>0$). Trying to closely reproduce their measurements on our models we obtain a ratio $(\sigma_r/\sigma_t)_{{\re}/4}\approx1.0$ for all five galaxies, with maximum differences of 10\% between ours and their results. This consistency test is not as useful as the previous one, as it does not allow us to test our derivation of the shape of the velocity ellipsoid in flattened galaxies. However for the two nearly round objects NGC~3379 and NGC~3608 both sets of models show that these galaxies are very nearly isotropic.

In summary, the comparison between our anisotropy determinations and those of previous authors, on two limited samples of 5 models each, indicates a general agreement and suggests that our systematics errors are on the order of 5\% on the shape of the velocity ellipsoid.

\subsection{Comparison with the \vse\ diagram}
\label{sec:vsigma}

In the previous sections we studied the anisotropy in galaxies by direct inversion of the observed kinematics and photometry, using axisymmetric dynamical models. In this section we revisit the previous findings with the classic \vse\ diagram \citep{bin78}, as was done for the last thirty years. This will provide a link between our results and the classic ones, and in addition it will allow us to test the robustness of the derived values.

\begin{figure}
    \plotone{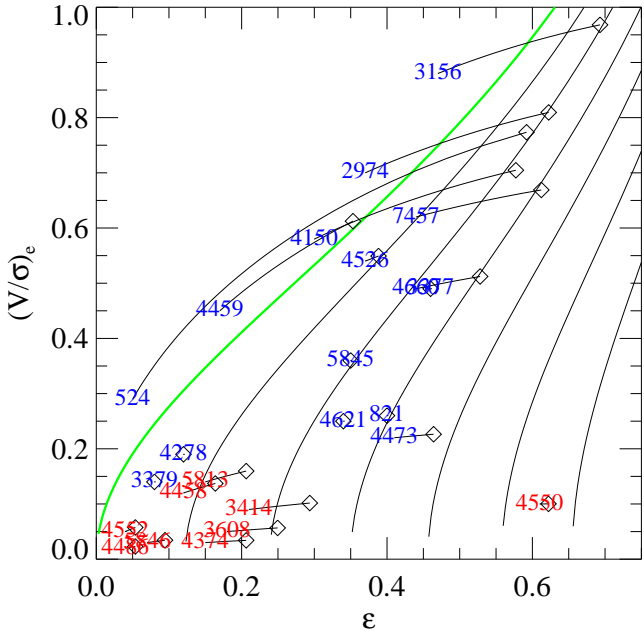}
    \caption{\vse\ diagram for our modelling subsample of 24 galaxies. The red and blue labels refer to the NGC number of the slow and fast rotators respectively, and show the measured values of the luminosity weighted ellipticity $\varepsilon$ and \vs. The solid lines, starting from each object show the effect of correcting the observed values of each galaxy to an edge-on view (diamonds). For the fast rotators we adopted the inclination derived from the models of Paper~IV. For the slow-rotators we show the correction for inclination assuming the average inclination of a randomly oriented sample $i=60^\circ$. Even under this assumption the corrections are small for the slow-rotators. The grid of solid curves shows the location on this diagram of edge-on oblate galaxies with different anisotropy $\delta=0,0.1,\ldots,0.6$ (edge-on isotropic models $\delta=0$ are shown with the thick green line).}
    \label{fig:v_over_sigma}
\end{figure}

In \reffig{fig:v_over_sigma} we visualise the location of the slow and fast rotators samples in the \vse\ diagram. Traditionally the observed \vs\ quantity is computed from the central velocity dispersion and the maximum rotational velocity \citep[but see][]{bac85}. Recently \citet{bin05} updated the formalism to compute the quantity in a more robust way, from integral-field data. Here the availability of \sauron\ integral-field kinematics allow us to apply this new formalism for the first time to real galaxies. We use the updated formulae and define
\begin{equation}
\left(\frac{V}{\sigma}\right)_e^2\equiv\frac{\langle V^2 \rangle}{\langle\sigma^2 \rangle} = \frac{\sum_{n=1}^{N} F_n\, V_n^2}{\sum_{n=1}^{N} F_n\, \sigma_n^2}
\label{eq:vsigma}
\end{equation}
as a luminosity-weighted quantity, which we estimate from the binned \sauron\ kinematics. Here $F_n$ is the flux contained inside the $n$-th Voronoi bin and $V_n$ and $\sigma_n$ the corresponding measured mean velocity and velocity dispersion. Similarly we define the ellipticity as
\begin{equation}
(1-\varepsilon)^2=q^2=\frac{\langle y^2 \rangle}{\langle x^2 \rangle} = \frac{\sum_{n=1}^{N} F_n\, y_n^2}{\sum_{n=1}^{N} F_n\, x_n^2},
\label{eq:eps}
\end{equation}
where the $(x,y)$ coordinates are centred on the galaxy nucleus and the $x$ axis is aligned with the galaxy photometric major axis. We estimate $\varepsilon$ from the individual \sauron\ pixels, inside a given galaxy isophote, within the same region used for the computation of \vs. This way of measuring $\varepsilon$ by diagonalizing the inertia tensor of the galaxy surface brightness is the standard technique among $N$-body modellers \citep[e.g.][]{car80} and for the automatic determination of $\varepsilon$ in large galaxy surveys \citep[e.g.][]{ber96}. When the ellipticity varies little with radius the $\varepsilon$ measured with this technique is very similar to the intensity-weighted $\varepsilon$ \citep{ryd99} from the standard photometry profiles (see also \reffig{fig:comparison_bender94_eps}). In general however the $\varepsilon$ determinations from the moments are biased towards larger radii than $\varepsilon$ determinations from the profiles.

For a rigorous application of the \vse\ diagram, the summation~(\ref{eq:vsigma}) should extend to infinite radii and the $\varepsilon$ should be constant with radius. In Appendix~\ref{sec:test_iso} we show that reliable results for the \vse\ values (at least in the limited isotropic case considered) can still be obtained when the summation is spatially limited and the ellipticity is slowly varying. To make the measurement comparable between different galaxies, we limit the summation to 1\re, or to the largest observed radius, whichever is smaller. We also show that better results are obtained when the summation is performed inside ellipses instead of circles. For this reason the quoted values of \vs\ are measured inside ellipses of area $A=\pi{\re}^2$ and semi-major axis $a=\re/\sqrt{1-\varepsilon}$, where the mean ellipticity $\varepsilon$ is measured inside an isophote of the same area $A$. When the isophote/ellipse are not fully contained within the observed \sauron\ field of view, we limit the summation to the largest isophote/ellipse contained within the field.
The \vse\ values are given in Table~\ref{tab2}.

The \vse\ diagram for the 24 galaxies of the modelling subsample is shown in \reffig{fig:v_over_sigma}. The solid curves in the \vse\ diagram represent the location of oblate edge-on models with different anisotropy $\delta$. Real galaxies are generally not seen edge-on, so a correction for the effect of inclination is in principle needed to correctly interpret the diagram. This usually cannot be done, as the inclination of early-type galaxies is not easy to determine. In our case however, we have an estimate for the inclination of our galaxies (\refsec{sec:models}), so we can correct the position of our objects on the diagram to the edge-on case. Although the inclination we use is assumption-dependent, we showed in Paper~IV that it generally provides a big improvement from a complete lack of knowledge of the inclination.

For a galaxy observed at an inclination $i$ (edge-on corresponding to $i=90^\circ$), with an ellipticity $\varepsilon$ and a given ratio  $(\vs)_{\rm obs}$, the values corrected to an edge-on view are \citep[\S 4.3]{bin87}:
\begin{equation}
\left(\frac{V}{\sigma}\right)_e = \left(\frac{V}{\sigma}\right)_{\rm obs} \frac{\sqrt{1-\delta\cos^2 i}}{\sin i},
\label{eq:corr_vs}
\end{equation}
\begin{equation}
\varepsilon_{\rm intr}=1-\sqrt{1+\varepsilon (\varepsilon-2)/\sin^2 i},
\label{eq:corr_eps}
\end{equation}
Starting from a given set of edge-on parameters the anisotropy under the oblate assumption is obtained as \citep{bin05}
\begin{equation}
\delta = 1 - \frac{1+(\vs)^2}{\left[1-\alpha(\vs)^2\right]\;\Omega(e)}
\label{eq:delta_vs}
\end{equation}
where
\begin{equation}
e=\sqrt{1-(1-\varepsilon_{\rm intr})^2},
\end{equation}
\begin{equation}
\Omega(e)=\frac{0.5\left(\arcsin e - e \sqrt{1-e^2}\right)}
{e\sqrt{1-e^2} - \left(1-e^2\right) \arcsin e},
\end{equation}
and $\alpha$ is a dimensionless number, which does not depend on the galaxy amount of rotation, but only on how the stellar density $\rho$ and streaming velocity $v_\phi$ are distributed in the $(R,z)$ plane. In Appendix~\ref{sec:test_iso} we use Jeans models to show that a value $\alpha\approx0.15$ places isotropic models, with a realistic surface brightness distribution, close to the isotropic line on the \vse\ diagram. Although $\alpha$ may in principle vary for different galaxies, tests suggest variations to be small and we will use this value for all the \vse\ diagrams in this paper.
As the inclination correction requires the knowledge of both the known inclination $i$ and the unknown anisotropy $\delta$, one has to proceed in an iterative manner. However this process converges very rapidly, as the correction depends very little on $\delta$. The effect of correcting the observed values of our galaxies on the \vse\ diagram is shown in \reffig{fig:v_over_sigma}.

The comparison between the anisotropy derived from the inclination-corrected \vse\ diagram and from the dynamical models is shown in the bottom panel of \reffig{fig:anisotropy}. The generally good agreement confirms the reliability of the two approaches and the significant anisotropy of most of the fast rotators in the modelling subsample.

As pointed out by \citet{kor82a}, in the formalism of \citet{bin78} the isotropic line in the \vse\ diagram is approximated to good accuracy by the expression $(\vs)^2\approx\varepsilon/(1-\varepsilon)$. Similarly, in the revised formalism for integral-field kinematics of \citet{bin05}, which we adopt in this paper, the minimax rational approximation of order (1,1) to the isotropy line (equation~[\ref{eq:delta_vs}] with $\delta=0$ and $\alpha=0$) is given by
\begin{equation}
\left(\frac{V}{\sigma}\right)_e\approx0.890\sqrt{\frac{\varepsilon}{1-\varepsilon}},
\end{equation}
which has a maximum relative error of just 0.3\% over the whole interval $\varepsilon=[0.01,0.99]$. For the adopted value of $\alpha=0.15$ (Appendix~\ref{sec:test_iso}) the corresponding approximation to the isotropic line becomes
\begin{equation}
\left(\frac{V}{\sigma}\right)_e\approx0.831\sqrt{\frac{\varepsilon}{1-0.896\,\varepsilon}},
\end{equation}
which has a maximum fractional error of just 0.2\% in the same interval.

\subsection{Understanding the anisotropy}
\label{sec:understand}

\begin{figure*}
	\includegraphics[width=0.75\textwidth]{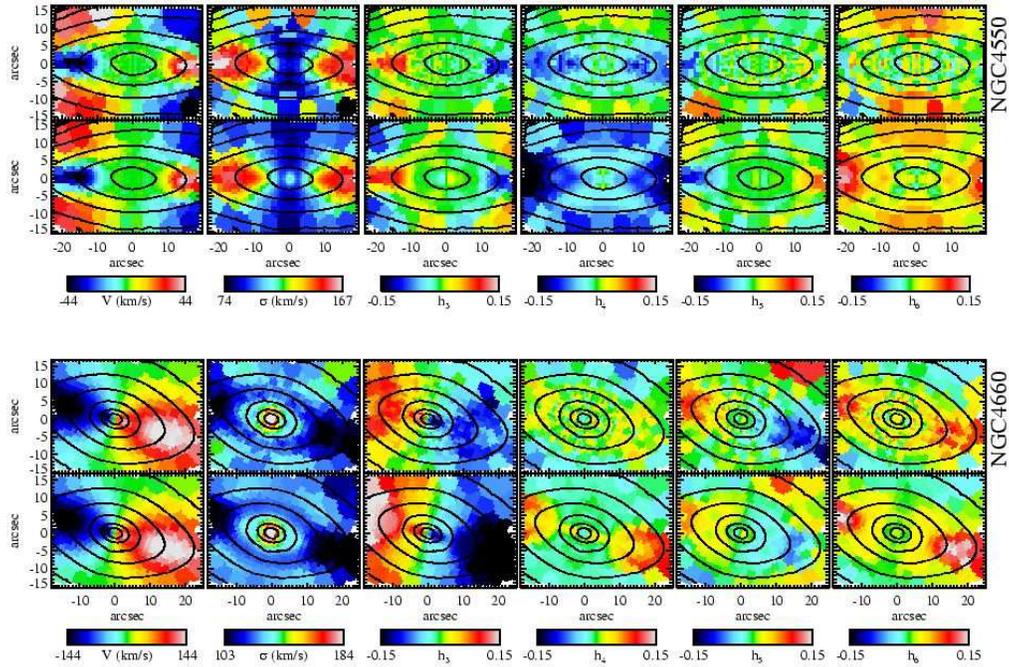}
    \caption{Data-model comparison. Each column from left to right shows the kinematical moments: mean velocity $V$, velocity dispersion $\sigma$, and higher Gauss-Hermite moments h$_3$--h$_6$. {\it Top row:} Voronoi binned \citep{cap03} and bi-symmetrised {\tt SAURON} observations. {\it Bottom row:} Three-integral Schwarzschild best fitting model. The two fits correspond to NGC~4550 (top) and NGC~4660 (bottom) respectively.}
    \label{fig:models}
\end{figure*}

We saw in \refsec{sec:global} that the anisotropy of early-type galaxies tends to be due to a flattening of the velocity ellipsoid along the $z$ direction, parallel to the symmetry axis. However two galaxies, NGC~4473 and NGC~4550, stand out for being dominated by tangential dispersion. Understanding what makes some object special is key to understand the normal galaxies.

\begin{figure}
    \plotone{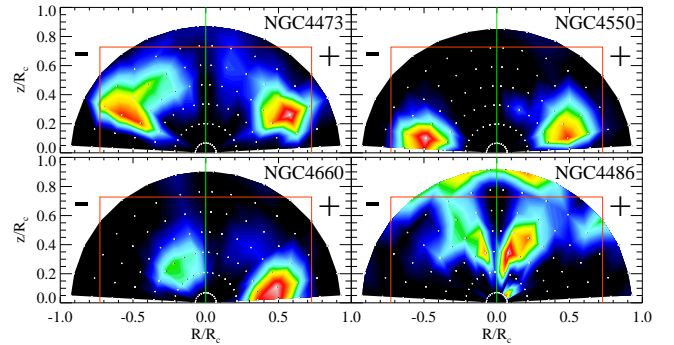}
    \caption{Integral space at a given energy for the solution of the Schwarzschild models for the galaxies NGC~4473, NGC~4550, NGC~4660 and NGC~4486. Each panel plots the meridional plane $(R,z)$ with the location (white dots) where orbits are started with $v_R = v_z = 0$ at the given energy. For nearly edge-on galaxies, the white dots also corresponds to the position of the orbital cusps, where every orbit gives its strongest contribution to the observables on the sky plane. The energy was chosen as that of a circular orbit with radius $R_c=32\arcsec$, which is about the size of the observed \sauron\ field (red rectangle). The coloured contours show the fraction of mass assigned to different orbits at that energy, where bright colours corresponds to high mass fractions. Orbits at negative and positive $R$ starting conditions correspond to prograde and retrograde respectively. Orbits with high angular momentum are found in the bottom right and bottom left corner respectively on the plots. Orbits near the symmetry axis (green line) have low angular momentum. Orbits near the equatorial plane ($z=0$) are intrinsically flat. See Fig.~6 of Paper~IV for a detailed explanation of this diagram.}
    \label{fig:integral_space}
\end{figure}

In \reffig{fig:models} we show the data versus model comparison for the galaxy NGC~4550, while the same comparison for NGC~4473 was shown in Fig.~2 of \citet{cmd05}. Along the major axis both galaxies show a peculiar decrease in the $V$ field and a corresponding increase in the $\sigma$ field. As already pointed out by \citet{rub92} and \citet{rix92}, for the case of NGC~4550, a natural explanation for this behaviour is the presence of two counterrotating stellar components, as indicated by the observed double-peaked line profiles. In \reffig{fig:integral_space} this explanation is confirmed from the solution of the Schwarzschild models, which shows two major kinematically distinct stellar components, rotating in opposite directions. In all the fits we adopted a modest regularization ($\Delta=10$; see \citet{van98} for a definition). In addition the \sauron\ data allow us to find that the two disks have very different scale-heights. Integrating the solution over the whole model we find that in NGC~4450 the counterrotating component constitutes about 50\% of the total mass, while in the case of NGC~4473 the counterrotating component is about 30\% of the total mass. A more detailed analysis of the incidence of kinematically distinct components in early-type galaxies goes beyond the scope of the present paper.

For comparison with the two peculiar cases, in \reffig{fig:models} we also show the best fit model for the prototypical fast rotating disky elliptical NGC~4660 \citep{ben94} and the corresponding model solution in \reffig{fig:integral_space}. We also show the model solution for the prototypical slow rotating giant elliptical NGC~4486 (M87; the best fit model was shown in Fig.~3 of \citealt{cmd05}). It appears that the fast rotators tend to be characterized by flat stellar components (orbital starting conditions close to the equatorial plane) with angular momentum significantly smaller than the circular orbit (starting conditions far from the bottom left/right corners of the plot). The relatively low angular momentum of these flat components produces a significant radial dispersion, which is the main reason of the observed anisotropy. The dominant radial dispersion in these flat disk-like stellar components is reminiscent of the one observed in disk galaxies, and generally attributed to disk heating processes \citep[e.g.][]{sch03}. In some cases the disk-like components may happen to counterrotate, or two disks may be present, in which case tangential anisotropy dominates the observed anisotropy.

Interestingly the disk-like components, which tend to characterise the fast rotators, not only seem to be distinct in integral space, but also differ in terms of their stellar population. As shown in Fig.~11 of \citet[Paper~VI]{kun06}, all the flat fast rotators are characterized by an Mgb line-strength distribution which is flatter than the isophotes. The fact that a flat Mgb is seen in all the flat galaxies suggests that perhaps {\em all} the fast rotators contain this metallicity-enhanced disk-like component, which is only visible in the flattest objects because they are closer to edge on. This result is reminiscent of the finding by \citet{lau05} that all flat cuspy galaxies show a disk. This metallicity enhancement indicates that additional star formation activity happened in the disk and, not surprisingly, implies that gas dissipation was involved in the formation of these flat structures. The radial anisotropy however shows that heating was significant after the disk formation, as otherwise the stars in the flat components would still move on orbits that are closer to circular.

\subsection{Relation of anisotropy with other global observables}
\label{sec:correlation}

\begin{figure}
    \plotone{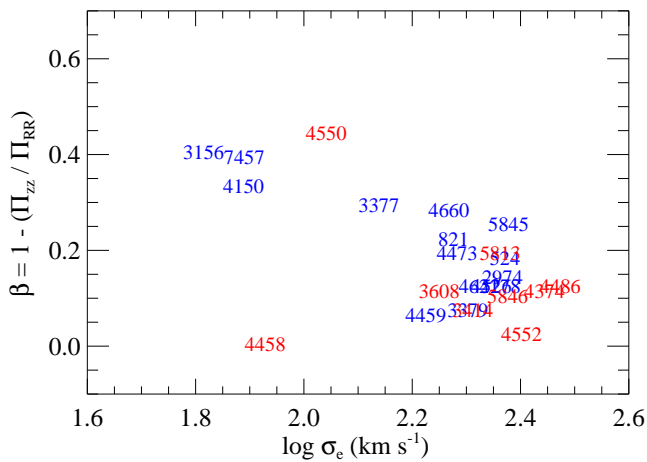}
    \caption{Anisotropy parameter $\beta$ in the meridional plane versus the luminosity-weighted second moment \se\ of the stellar velocity (from Paper~IV). The red and blue labels refer to the slow and fast rotators respectively.}
    \label{fig:sigma_beta}
\end{figure}

In \reffig{fig:sigma_beta} we show the correlation between the anisotropy and the galaxy velocity dispersion \se\ within the effective radius, taken from Paper~IV. There is a trend for the most massive galaxies to have a {\em smaller} anisotropy within one \re, with the exception of the special case NGC~4458 (see also Paper~IV, \S~5.1). In this diagram we use the parameter $\beta$ to characterise the anisotropy. This parameter is measured in the galaxy meridional plane so it describes the orbital distribution in a way that is not affected by the direction of rotation of the stars in the galaxy. The $\beta$ parameter measures the same anisotropy e.g.\ in two galaxies that formed with the same physical process, but in which one galaxy experienced a merger in a prograde direction and the other in a retrograde direction.

\begin{figure}
    \plotone{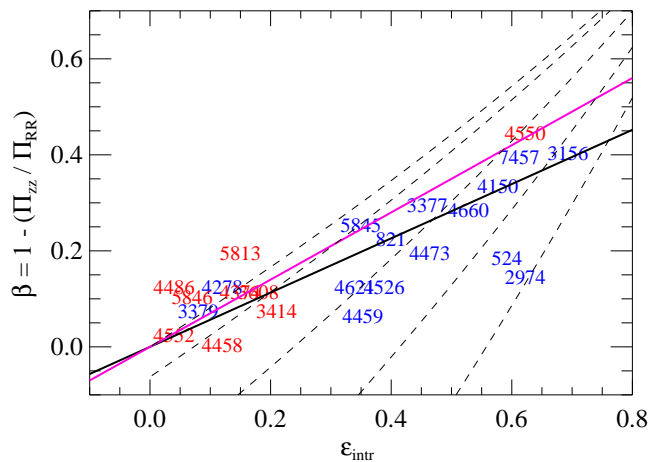}
    \caption{Anisotropy parameter $\beta$ in the meridional plane versus the average {\em intrinsic} ellipticity $\varepsilon_{\rm intr}$ of the galaxies. The observed anisotropy is related to the intrinsic ellipticity of the galaxies. The thick black solid line is the best fitting relation. The dashed lines show the predictions of the tensor virial equations, in the form of equation~(\ref{eq:delta_vs}), for different values of the $\vs=0,0.25,\ldots,1$ (see text for details). The red and blue labels refer to the slow and fast rotators respectively. The magenta line show the same relation $\beta=0.7\varepsilon_{\rm intr}$ as the one in \reffig{fig:v_over_sigma_lines}.}
    \label{fig:epsilon_beta}
\end{figure}

A general trend is found between anisotropy and the intrinsic galaxy ellipticity (\reffig{fig:epsilon_beta}). The best-fitting relation to this rather scattered distribution of points, using a robust bisector algorithm, has the form:
\begin{equation}
    \beta=(0.6\pm0.1)\,\varepsilon_{\rm intr}.
    \label{eq:beta_corr}
\end{equation}
Given that the errors on both $\beta$ and $\varepsilon_{\rm intr}$ are model dependent and difficult to estimate, we adopted constant errors on both variables. The quoted error on the slope was determined by enforcing the condition $\chi^2=\nu$, where $\nu$ is the number of degrees of freedom of the fit.

A better understanding of the meaning of the observed trend comes by considering the region in the $(\beta,\varepsilon_{\rm intr})$ plane that is allowed by the tensor virial equations. Equation~(\ref{eq:delta_vs}) defines in fact a relation between the three variables $\varepsilon_{\rm intr}$, \vs\ and $\delta$. This equation is usually visualised in the observational plane $(\vs,\varepsilon_{\rm intr})$, for different values of $\delta$ as in \reffig{fig:v_over_sigma}. The same equation can be plotted in the theoretical plane $(\delta,\varepsilon_{\rm intr})$, for different values of \vs.  As we saw that in real galaxies we have in general $\beta\sim\delta$ (\reffig{fig:anisotropy}), we will use the relation to understand the location of galaxies in the $(\beta,\varepsilon_{\rm intr})$ plane (\reffig{fig:epsilon_beta}). As expected, within the numerical approximations and the limited spatial coverage, the galaxies generally lie within the region allowed by the tensor virial equations. The observed trend implies a specific form of the increase of the \vs\ with increasing $\varepsilon_{\rm intr}$.

A possible caveat is that a trend of the {\em average} anisotropy with increasing $\varepsilon_{\rm intr}$ should be expected even if the galaxies were uniformly distributed in $\beta$ at every $\varepsilon_{\rm intr}$, due to the limits on $\beta$ imposed by the virial equilibrium. However the 24 galaxies in the modelling subsample do not appear uniformly distributed in $\beta$. Moreover we show in the Appendix~\ref{sec:simulation} that a uniform distribution in $\beta$ is not able to reproduce the observed distribution of all the galaxies of the \sauron\ sample on the \vse\ diagram: a general trend, and an upper limit of anisotropy as a function of ellipticity is required to explain the data. As this trend is not implied by the tensor virial equations alone (nothing prevents all galaxies to scatter around $\beta\sim0$), the observed distribution must hold a fossil record of the process that led to the galaxy formation.

One explanation of the correlation may be that flatter galaxies were more strongly affected by dissipation before star formation was completed. Dissipation naturally makes a galaxy flat by reducing the kinetic energy in the direction parallel to the global angular momentum. Dissipation is also required to produce the small observed kinematical misalignments for the fast rotators of the whole sample (\reffig{fig:kinematics_pa_48}).

\section{\vse\ diagram of the SAURON sample}
\label{sec:vs_sauron}

In \refsec{sec:modeling} we used dynamical modelling to recover the anisotropy of a sample of 24 galaxies extracted from the \sauron\ representative sample. The galaxies in our sample were separated into two groups of fast and slow rotators (Paper~IX). The slow rotators all appear to have small anisotropy, while the anisotropy of the fast rotators seem to be related to the intrinsic ellipticity of each galaxy. One limitation of the analysis is due to the fact that the sample for which dynamical models were constructed is not necessarily representative of the whole galaxy population. To address this problem we show here that the findings of \refsec{sec:modeling} can be extended to the whole \sauron\ sample of 48 E/S0 galaxies. The results still holds even when we include in the analysis 18 additional E/S0 galaxies that are not part of the survey, but for which \sauron\ data also exist.

\subsection{Intrinsic shape distribution}
\label{sec:shape}

A first point to address, to be able to simulate the observed distribution of galaxies in the \vse\ diagram, is to understand what is the shape distribution of the galaxies in the \sauron\ sample. This is in principle a very difficult and intrinsically degenerate task, as even the full knowledge of the observed distribution of galaxy shapes and of their projected kinematical misalignments are not sufficient to recover the corresponding intrinsic quantities, unless strong assumptions are made \citep{fra91}. However the availability of integral field stellar kinematics allows us to realise that the situation is simpler for real galaxies and that strong conclusions can be reached without any statistical analysis.

\begin{figure}
    \plotone{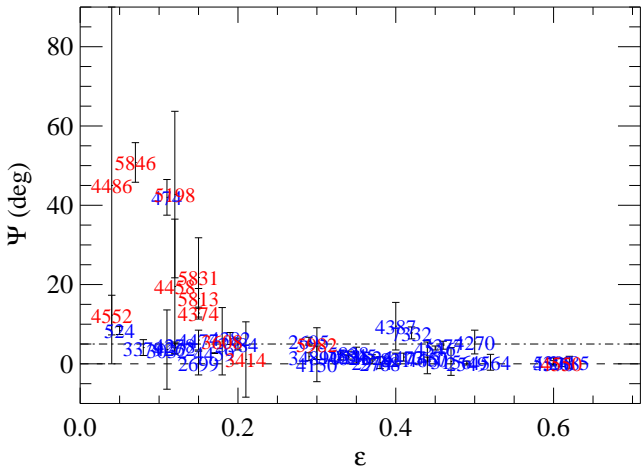}
    \caption{Kinematical misalignment $\Psi$ between the global photometric PA and the mean kinematic PA on the \sauron\ field. The red and blue labels refer to the slow and fast rotators respectively. We find that (i) all galaxies flatter than $\varepsilon\sim0.2$ have $\Psi\sim0$, and (ii) all the fast rotators have $\Psi\sim0$, for any $\varepsilon$ (exception is the recent merger galaxy NGC~474). Assuming the \sauron\ survey constitutes a representative sample, this implies that triaxial galaxies have $(c/a)\ga0.8$. Strongly prolate galaxies are also excluded from this sample.}
    \label{fig:kinematics_pa_48}
\end{figure}

In \reffig{fig:kinematics_pa_48} we plot the kinematical misalignment $\Psi$ for the whole sample of 48 E/S0 galaxies. $\Psi$ is defined as the angle between the projected minor axis of the surface brightness and the direction of the projected kinematical rotation axis, which corresponds to the projected direction of the intrinsic angular momentum \citep{fra91}:
\begin{equation}
 \sin\Psi=|\sin(\rm PA_{kin}-PA_{phot})|\qquad 0^\circ\le\Psi\le90^\circ.
\end{equation}
The $\rm PA_{kin}$ is defined as the PA at which the observed \sauron\ velocity field is best approximated by a bi-antisymmetric version of it and was measured as described in Appendix~C of \citet{kra06}.

The position angle (PA) of the photometric major axis is obtained using the moments of the surface-brightness \citep{car80} from the large scale MDM photometry (Falcon-Barroso et al.\ in preparation). It generally agrees well with the global PA quoted by catalogues like RC3 and the 2MASS Extended Source Catalog \citep{jar00} when available.
It is important to consider a global large-scale PA because: (i) early type galaxies tend to be rounder in their central regions than in the outer parts \citep{ryd01}. Thus the photometric axes are generally more accurately determined at large radii. (ii) The central regions can be easily affected by bar perturbations. In the case of a disk galaxy with a bar we are not interested in the PA of the bar, but we want to measure the PA of the main disk, which is better measured at large radii \citep[e.g.][]{erw03}. In most cases however, a constant-PA surface brightness distribution describes well the whole galaxy and no distinction needs to be made.

It turns out that nearly all the fast rotators in \reffig{fig:kinematics_pa_48} have a misalignment $\Psi\sim0^\circ$ within the measurement errors. In the few cases where the misalignment reaches $\Psi\la5^\circ$ (e.~g.\ NGC~3377, NGC~3384, NGC~4382, NGC~7332) the nonzero value can be ascribed to the presence of a bar. The only exception is the galaxy NGC~474 for which $\Psi\approx40^\circ$. This galaxy appears distorted by a recent merger, as evidenced by prominent stellar shells at large radii \citep{tur99}. Interestingly, also for the obvious nearly face-on bars NGC~4477 and NGC~4262, characterized by the typical oval shape of the isophotes (see Paper~III), the kinematical PA appears not affected by the bar and is still well aligned with the outer disk, which defines the photometric PA.

In a stationary triaxial system kinematical misalignments are common due to two reasons: (i) the intrinsic angular momentum can lie anywhere in the plane containing the long and short axis, so the kinematical axis are not necessarily aligned with the principal axes of the density; and (ii) projection effects can produce observed misalignments even when the angular momentum is aligned with a principal axis. The fact that $\Psi\sim0$ for all the fast rotators can then only be interpreted as strong evidence that these systems are nearly oblate systems. In some cases the fast rotators contain bars, which provide further evidence for their disk-like nature.

The situation is completely different for the slow rotators. All these objects appear quite round in projection, the flattest one having $\varepsilon\la0.3$ (excluding the special case NGC~4550). Some of these galaxies show significant kinematical misalignments, although in general the rotation is confined only to the central region, so it is not representative of the whole galaxy. But some galaxies like NGC~5198 can only be genuine triaxial systems (see Paper~III). However the maximum observed ellipticity sets a limit of $c/a\ga0.7$ on the ratio between the shortest and longest axis of the density. A complete larger sample of slow rotating galaxies or detailed triaxial dynamical modelling would be needed to investigate the shape distribution of these objects in more detail. The observed difference in shape between the fast rotating galaxies and the generally brighter slow rotating ones is consistent with the bimodality in the shape distribution between the fainter and brighter galaxies observed by \citet{tre96}. The availability of the stellar kinematics however shows that the shape distribution, not surprisingly, is more closely related to the kinematics than to the luminosity alone.

\subsection{Projection effects on the \vse\ diagram}
\label{sec:vs_lines}

We established in Paper~IX and further verified in \refsec{sec:shape} that the whole \sauron\ sample of E/S0 galaxies can be broadly subdivided into a class of fast rotating and nearly oblate systems and another class of weakly triaxial slowly rotating galaxies.
In \refsec{sec:correlation} we saw that the anisotropy $\beta$, measured in the meridional plane of the 24 galaxies in the modelling subsample, is roughly described by the relation $\beta\sim0.6\,\varepsilon_{\rm intr}$. Moreover in \refsec{sec:global} we found that with a few notable exceptions $\beta\sim\delta$, where $\delta$ is the anisotropy as can be inferred from the \vse\ diagram. This implies that in most cases, one should also have $\delta\sim0.6\,\varepsilon_{\rm intr}$. If these relations that we found for a small sample are valid in general, we should be able to model in a statistical way the distribution of the whole sample of fast rotators in the \vse\ diagram, as a random sample of oblate galaxies, with anisotropy defined by their intrinsic ellipticity $\varepsilon_{\rm intr}$. The slow rotators should be inconsistent with this distribution, as they are not expected to be well described by oblate systems.  A rigorous test of this idea is not possible due to the complex nature of our sample selection effects (\refsec{sec:selection}), however an attempt is still performed in Appendix~\ref{sec:simulation} via Monte Carlo simulation. There we show that the observed distribution of galaxies on the \vse\ diagram is indeed consistent with a nearly linear trend of anisotropy with ellipticity.

\begin{figure}
    \plotone{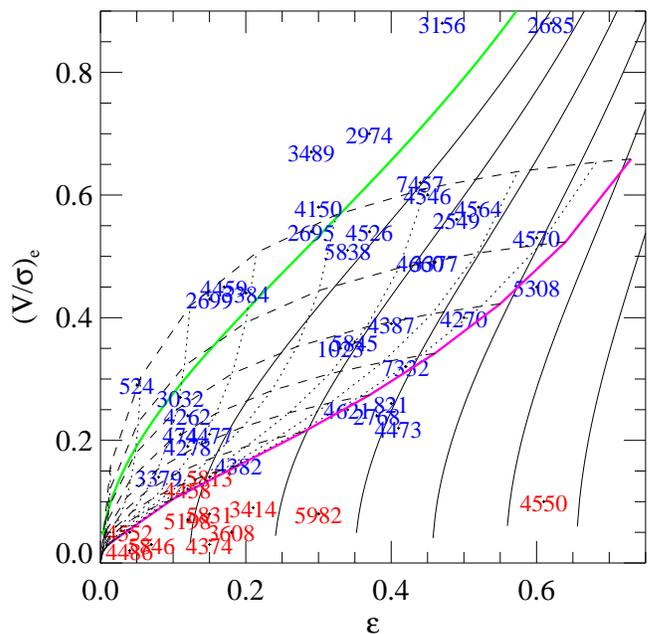}
    \caption{\vse\ diagram for the 48 galaxies in the \sauron\ representative sample. The red and blue labels refer to the slow and fast rotators respectively (Table~\ref{tab2}). The magenta line corresponds to the relation $\delta=0.7\,\varepsilon_{\rm intr}$ for edge-on galaxies. The dotted lines show the location of galaxies, originally on the magenta line, when the inclination is varied. Different lines are separated by steps of 10$^\circ$ in the inclination. The dashed lines are equally spaced in the intrinsic ellipticity.}
    \label{fig:v_over_sigma_lines}
\end{figure}

Here we just try to obtain a qualitative understanding of the effect of the inclination on the \vse\ diagram. For this we plot in \reffig{fig:v_over_sigma_lines} with a magenta line\footnote{A compact rational approximation of the magenta line which appears in the \vse\ diagram of \reffig{fig:v_over_sigma_lines} is given by:
\begin{equation}
\left(\frac{V}{\sigma}\right)_e\approx\sqrt{\frac{(0.09+0.1\,\varepsilon)\;\varepsilon}{1-\varepsilon}},
\end{equation}
which has a 1\% accuracy in the useful interval $\varepsilon=[0.01,0.7]$.} a linear relation\footnote{Any relation between anisotropy and ellipticity for oblate galaxies has to satisfy the necessary condition $(\vs)^2\ge0$ in equation~[\ref{eq:delta_vs}], or by Taylor expansion
\begin{equation}
\delta\le1-1/\Omega(\varepsilon)\approx0.8\varepsilon+0.15\varepsilon^2+0.04\varepsilon^3+\ldots
\end{equation}
This implies that the steepest linear relation is $\delta=0.8\,\varepsilon$, independently of $\alpha$. The magenta line is less steep than this limit.} $\delta=0.7\,\varepsilon_{\rm intr}$  for edge-on galaxies.
This relation approximately traces the lower envelope described by the location of the observed fast-rotating galaxies on the \vse\ diagram. It is steeper than the formally best-fitting relation~(\ref{eq:beta_corr}), but still within the large errors. It is useful to have a qualitative understanding of the projection effects, as the paths followed by galaxies of different intrinsic $(\delta,\varepsilon_{\rm intr})$ are nearly parallel, when the inclination is varied (e.g.\ \reffig{fig:v_over_sigma}). For a given intrinsic ellipticity $\varepsilon_{\rm intr}$ and anisotropy $\delta$, the corresponding \vs\ value, for an edge-on view is computed with the inverse of \refeq{eq:delta_vs}.
The dotted lines in \reffig{fig:v_over_sigma_lines} show the location of the galaxies, originally edge-on on the magenta line, when the inclination $i$ is decreased. The projected $V/\sigma$ and ellipticity values at every inclination are computed with the inverse of equations~(\ref{eq:corr_vs}, \ref{eq:corr_eps}).

It appears that indeed all the fast-rotators lie on the \vse\ diagram to the left of the magenta line of the edge-on galaxies, as expected if they were a family of oblate models following a general trend of increasing $\delta$ for larger $\varepsilon$. The upper envelope of the region spanned by the lines of different inclinations is also generally consistent with the location of the observed galaxies, with reasonable values of the maximum intrinsic ellipticity and anisotropy. In general, the fast-rotators for which the inclination was determined independently lie near a similar inclination on the diagram.
The slow rotators on the other hand do not lie in the region allowed by oblate models following the given $\delta-\varepsilon$ relation. They may not follow any such relation or, more likely, they may not be oblate. This is consistent with the result of \refsec{sec:shape} that the fast and slow rotating galaxies cannot be described by a single shape distribution. In Appendix~\ref{sec:simulation} we also discover that a sharp truncation below the magenta seems required to explain the observations. This zone of avoidance in the \vse\ diagram may be related to the stability of axisymmetric bodies with high ellipticity and large anisotropy, and needs to be confirmed with a larger and unbiased sample of galaxies.

Remarkably, only two galaxies fall significantly below the magenta line for $\varepsilon\ga0.3$. These are the fast-rotator NGC~4473 (marginally) and the special slow-rotator NGC~4550. This was expected from the fact that these two galaxies are dominated by tangential anisotropy (\reffig{fig:anisotropy}), due to the presence of two counterrotating disks (\reffig{fig:integral_space}). Although they follow the $\beta-\varepsilon$ relation, they do not satisfy the $\delta-\varepsilon$ relation as they have $\beta<\delta$.

Other significant exceptions are NGC~3156 and NGC~2685, which lie well above the general envelope described by the other galaxies, apparently next to the oblate isotropic line. Do these galaxies actually rotate as fast as an isotropic rotator? They are almost completely disk-dominated, as evidenced by the nearly constant velocity dispersion maps, without clear signature of a hot central stellar component (Paper~III). If this is the case they must be intrinsically quite flat and rotate significantly less than an isotropic rotator. For NGC~3156 the inclination $i\approx68^\circ$ determined via modelling in Paper~IV agrees with the inclination inferred from the shape of the outer isophotes, under the assumption of a thin disk geometry. At that inclination the models show that the galaxy indeed follows the general trend of anisotropy observed for the other galaxies (\reffig{fig:epsilon_beta}). A similar discussion is likely to apply to NGC~2685. These two galaxies appear to be special cases of flat and disk-dominated galaxies which are poorly represented in our sample of early-type galaxies.

The observed anisotropy of the fast-rotators is consistent with the finding by \citet{bur05} who recently pointed out that the observed distribution of disky ellipticals on the \vse\ diagram does not imply isotropy, but can be explained by projection effects on a family of anisotropic and nearly oblate merger remnants, as they obtained as a result of $N$-body simulations.

\subsection{E and S0 galaxies on the \vse\ diagram}
\label{sec:bulges}

In Paper~IX we showed that early-type galaxies can be subdivided into two homogeneous classes of slow and fast-rotators respectively. The first class is composed almost exclusively of E galaxies, while the second class is composed both of E and S0 galaxies. Now S0 galaxies are defined by the presence of a disk component and it is natural to ask whether all the fast rotators contain a disk, which is not always visible photometrically due to projection effects \citep{rix90}. In fact the disk becomes more difficult to detect when the galaxy is away from edge-on. To try to answer this question we can analyse the galaxies that, according to our simple explanation of \refsec{sec:vs_lines} are supposed to be close to edge-on. If they are indeed edge-on and if they all contain a disk component, this should be visible in the photometry. A detailed analysis of the photometric data will appear in a future paper in this series (Falcon-Barroso et al. in preparation) but we present here a qualitative discussion in relation to the \vse\ diagram.

We find a range of properties in galaxies near the edge-on magenta line in \reffig{fig:v_over_sigma_lines}. At the lowest $\vs\sim0.1$ there is the S0 galaxy NGC~4382, which shows evidence of interaction, stellar shells and distorted photometry at large radii from our MDM photometry. At higher  $\vs\sim0.2$ we have NGC~821, NGC~4473 and NGC~4621, which are classified as E (Table~\ref{tab2}), but show strongly disky isophotes and would be classified as disky ellipticals E(d) according to \citet{kor96}. An exception is NGC~2768, which does not have evidence for a disk. At higher $\vs\ga0.3$ we find only S0 galaxies. NGC~4270 has a peanut-shaped bulge, indicative of a nearly edge-on bar, NGC~3377 has strongly disky isophotes, while NGC~4570 and NGC~5308 have a clear thin disk component.

\begin{figure}
    \plotone{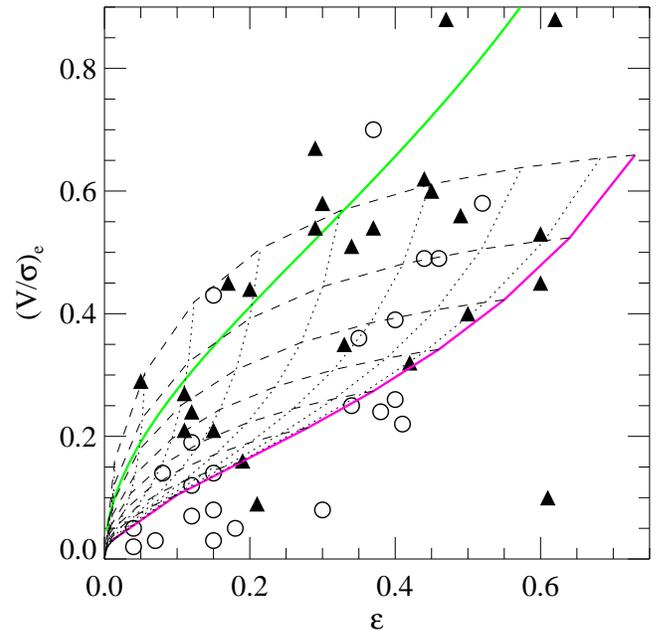}
    \caption{Galaxy morphology on the \vse\ diagram. Same as \reffig{fig:v_over_sigma_lines} but here filled triangles represent S0 galaxies and open circles Es. The envelope of the fast-rotators is populated mainly by S0 in the upper part and by E in the lower part.}
    \label{fig:vsigma_morphology}
\end{figure}

Although the current galaxy sample is too small to reach firm conclusions, we speculate that the observed trend in galaxy properties along the edge-on magenta line can be generalized to understand the whole distribution of the fast-rotators on the \vs\ diagram. The magenta line seems to describe a sequence of decreasing bulge over total-light ratio $B/T$. The region of low \vs\ is populated by galaxies with $B/T\sim1$. At increasing \vs\ the edge-on magenta line is dominated by S0 galaxies, with $B/T\sim0.5-0.6$ \citep{sim86}. When inclination effects are included on the magenta line one should expect the upper part of the envelope of the fast-rotators to be still dominated by S0 galaxies, where the disk can still be detected even for a non edge-on view, while the lower part will be populated by non-disky E, where no evidence of the faint disk can be detected unless the galaxies are edge-on. This is what one can qualitatively infer from \reffig{fig:vsigma_morphology} where we show the same sample as in \reffig{fig:v_over_sigma_lines}, but with the morphological classification included. The decrease of $B/T$ along the magenta line may also be the reason for the finding of Paper~IX (their Fig.~7) that the fast-rotators with the largest $\lambda_{\re}$ (and thus generally the largest \vs) tend to be the least massive.

The fact that the fast-rotators contain disks constitute a problem for the characterization of their properties using the \vse\ diagram. The diagram in fact is rigorously valid only for galaxies with a density stratified on similar oblate ellipsoids. This condition is clearly not satisfied for galaxies containing a rounder bulge and a flatter disk component. It would be interesting to understand in detail the orbital distribution of the different photometric components in early-type galaxies, as this constitutes a powerful record of the formation history. This would also be in principle feasible with the integral-field data and Schwarzschild models we have on a few carefully chosen galaxies, as the models do not make any assumption about the galaxy being made by a single stellar component (see e.g.\ \reffig{fig:integral_space}). This goes beyond the scope of the present paper. Here we notice that, from the dynamical models, there seems to be a tendency for the anisotropy to be stronger near the equatorial plane of the few disk-dominated galaxies (\refsec{sec:momemnts}). If this is true for the fast-rotators in general, one may speculate that the relation between anisotropy and ellipticity is related to the increase of both quantities with increasing $B/T$. However the anisotropy is in general not confined to the disks alone, indicating the situation is more complex.

In this paper we do not attempt a disk-bulge decomposition as we focus on the statistical analysis of a large sample. We only consider the global properties of the stellar orbital distribution in the (possibly heterogeneous) region of the E and S0 galaxies enclosed inside $\sim1\re$. Our goal is to study the ordered nature of early-type galaxies as seen on the \vse\ diagram, in much the same way as a similar set of simple measurements of integrated properties of a heterogeneous set of E and S0 galaxies allows  the Fundamental Plane to be studied \citep[e.g.][]{jor96}.

\subsection{SAURON \vse\ diagram including 18 `specials'}
\label{sec:specials}

\begin{table*}
\caption{Measured parameters for the 18 E/S0 which are {\em not} part of the \sauron\ sample.}
\tabcolsep=3pt
\begin{tabular}{llcccccccccccc}
\hline
Galaxy Name & Type & T & \re\ & $R_{\rm max}/\re$ & $\varepsilon$ & $\sqrt{\langle V^2 \rangle}$ & $\sqrt{\langle\sigma^2 \rangle}$ & $(\vs)_{\rm e}$ & PA$_{\rm phot}$ & PA$_{\rm kin}$ & $\Delta$PA$_{\rm kin}$ & $\lambda_{\re}$ & Rotator   \\
            &      &   & (arcsec)   &                         &                     &  \kms                        & \kms                             &                 & (deg)      & (deg) & (deg)  &   & \\
      (1)  & (2)           & (3)   & (4)     & (5)    & (6)   & (7)   & (8)    & (9)    & (10)   & (11) & (12) & (13) & (14) \\
\hline
 NGC~221   &   cE2      & -4.7 &  30 & 0.87 & 0.24            &  23  &   49  &  0.47 &  161.1 & 163.0 &  2.2 & 0.38 & F \\
 NGC~720   &   E5       & -4.8 &  36 & 0.75 & 0.43            &  24  &  234  &  0.10 &  142.3 & 141.5 &  2.7 & 0.12 & F \\
NGC~1700   &   E4       & -4.7 &  18 & 1.14 & 0.27            &  46  &  226  &  0.20 &   89.6 &  88.0 &  1.0 & 0.22 & F \\
NGC~2320   &   E        & -4.8 &  37 & 0.55 & 0.37            &  95  &  278  &  0.34 &  139.0 & 141.5 &  1.2 & 0.34 & F \\
NGC~2679   &   SB0:     & -2.0 &  28 & 0.77 & 0.19$^\dagger$  &  41  &   99  &  0.42 &  165.0 & 140.0 & 10.0 & 0.33 & F \\
NGC~4168   &   E2       & -4.8 &  29 & 0.70 & 0.20            &  11  &  184  &  0.06 &  125.2 &  60.5 & 12.2 & 0.05 & S \\
NGC~4261   &   E2-3     & -4.8 &  36 & 0.56 & 0.26            &  28  &  302  &  0.09 &  158.0 &  56.0 &  2.7 & 0.09 & S \\
NGC~4365   &   E3       & -4.8 &  50 & 0.51 & 0.25            &  27  &  240  &  0.11 &   44.0 & 145.0 &  2.0 & 0.09 & S \\
NGC~4379   &   S0-pec:  & -2.8 &  17 & 1.18 & 0.26            &  37  &   95  &  0.39 &  100.0 & 103.5 &  3.2 & 0.40 & F \\
NGC~4406   &   S0(3)/E3 & -4.7 & 104 & 0.25 & 0.21            &  19  &  216  &  0.09 &  121.2 & 15.00 &  5.5 & 0.06 & S \\
NGC~4472   &   E2/S0    & -4.8 & 104 & 0.27 & 0.16            &  25  &  300  &  0.08 &  159.2 & 159.0 &  3.0 & 0.09 & S \\
NGC~4478   &   E2       & -4.8 &  13 & 1.56 & 0.24            &  49  &  157  &  0.32 &  152.0 & 155.5 &  0.5 & 0.27 & F \\
NGC~4551   &   E:       & -4.8 &  13 & 1.56 & 0.28            &  29  &   99  &  0.29 &   70.2 &  70.0 &  4.7 & 0.27 & F \\
NGC~4649   &   E2       & -4.6 &  69 & 0.39 & 0.15            &  38  &  318  &  0.12 &  100.5 &  97.0 &  0.5 & 0.13 & F \\
NGC~5866   &   S0$\_$3  & -1.2 &  40 & 0.50 & 0.47            &  50  &  149  &  0.34 &  127.6 & 126.0 &  0.5 & 0.28 & F \\
NGC~6547   &   S0       & -1.3 &  13 & 1.60 & 0.65            &  88  &  141  &  0.62 &  132.8 & 132.5 &  1.7 & 0.60 & F \\
NGC~6548   &   S?       & -1.9 &  29 & 0.81 & 0.12$^\dagger$  &  47  &  136  &  0.35 &  64.0  &  66.0 &  2.2 & 0.36 & F \\
NGC~7280   & SAB(r)0$+$ & -1.2 &  19 & 1.07 & 0.40            &  54  &   84  &  0.64 &  76.9  &  77.0 &  4.2 & 0.60 & F \\
\hline
\end{tabular}
\begin{minipage}{17.8cm}
Notes: The meaning of the various columns is the same as in Table~\ref{tab2}. Here the effective radii \re\ are mostly taken from RC3. Exceptions are NGC~221 for which \re\ was taken from Paper~IV, and NGC~2679, NGC~6547 and NGC~6548, for which \re\ was determined from our MDM photometry.

$^\dagger$ These two galaxies show nearly face-on bars. The ellipticity represents the one of the outer disk, which also defines the photometric PA.
\end{minipage}
\label{tab3}
\end{table*}

\begin{figure}
    \plotone{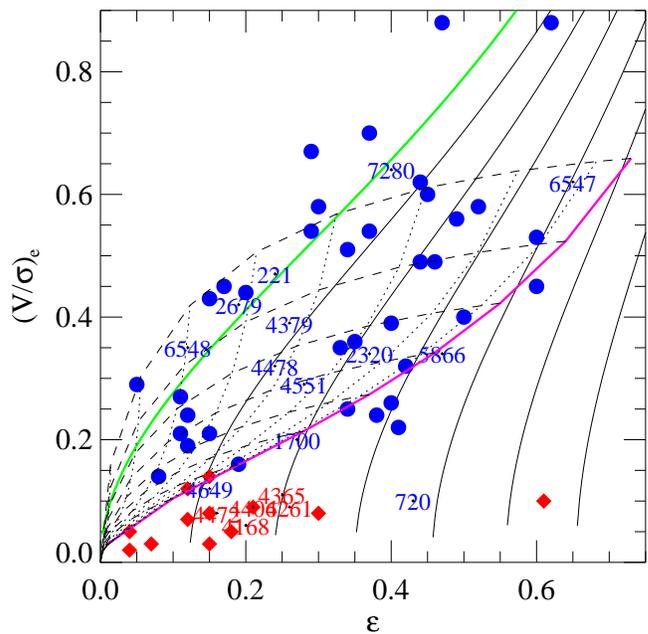}
    \caption{Same as \reffig{fig:v_over_sigma_lines} but for the 66 E/S0 galaxies observed with \sauron\ (48 galaxies from the main survey plus 18 specials). The red and blue labels refer to the slow and fast rotators respectively. The filled symbols refer to the galaxies in the \sauron\ survey, while the NGC numbers indicate the 18 `special' objects.}
    \label{fig:v_over_sigma_66}
\end{figure}

In addition to the 48 E/S0 galaxies of the \sauron\ representative sample presented in \refsec{sec:vs_lines}, we observed 18 additional E/S0 galaxies with \sauron, using the same setup as for the main survey. For these galaxies we extracted the stellar kinematics and measured the \vse\ and $\lambda_R$ using the same procedures and definitions as for the main sample (\refsec{sec:vsigma}) and give the numerical values in Table~\ref{tab3}. In \reffig{fig:v_over_sigma_66} we show an extension of \reffig{fig:v_over_sigma_lines}, where, in addition to the 48 galaxies of the \sauron\ survey, we also include the 18 specials. The new galaxies still populate the same regions as the galaxies from the main survey, confirming the result from that smaller sample. In particular all the fast-rotators still lie within the envelope defined by the $\delta-\varepsilon_{\rm int}$ relation (the magenta line) and its variation with inclination, while the slow-rotators lie in the bottom-left part of the \vse\ diagram. The five new slow-rotators are on average flatter than the ones in the survey sample, but still fall within the same ranges of low ellipticity ($\varepsilon\la0.3$).

A notable exception is NGC~720 which is at the edge of being classified as a fast-rotator (see Table~\ref{tab3}), but is quite flat and lies well below the magenta line. This galaxy shows little rotation, but displays a clear and symmetric velocity pattern, with no kinematical misalignment and no sign of photometric twist \citep{pel90}. All this indeed supports the classification of this galaxy as a fast-rotator and would suggest that the slow rotation may be due to counterrotating stellar components, as in the case of NGC~4473 and NGC~4550 (\refsec{sec:understand}). However, contrary to this hypothesis is the lack in NGC~720 of an elongation of the $\sigma$ field along the galaxy major axis. Two-integral dynamical models using the Jeans equations \citep{bin90} already showed that part of the slow rotation in this galaxy must be in the form of a flattened velocity ellipsoid in the meridional plane [$\beta>0$ i.e.\ DF not of the form $f(E,L_z)$]. This galaxy shows evidence from X-ray observations of being embedded in a triaxial dark-matter halo \citep{rom98,buo02} and the stellar component may be triaxial as well.
Detailed three-integral Schwarzschild's dynamical modelling is needed to clarify the nature of the orbital distribution in this interesting galaxy.

\begin{figure}
    \plotone{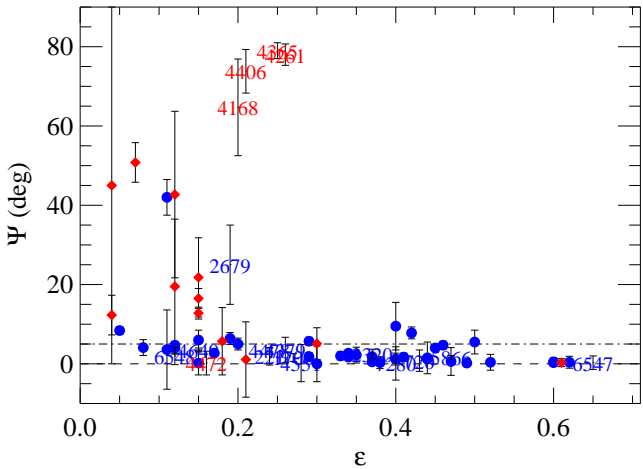}
    \caption{Same as \reffig{fig:kinematics_pa_48} but for the whole sample of 66 galaxies with existing \sauron\ stellar kinematics. Blue and red symbols refer to the fast/slow-rotators respectively. The filled symbols refer to the galaxies in the \sauron\ survey, while the NGC numbers indicate the 18 `special' objects. The morphology of NGC~2679 is dominated by a strong face-on bar, which explain the observed large misalignment. All the other new fast-rotators are consistent with a family of nearly oblate bodies and show no significant kinematical misalignment.}
    \label{fig:kinematics_pa_66}
\end{figure}

A key additional test provided by this extended sample concerns the shape distribution that we discussed in \refsec{sec:shape}. In \reffig{fig:kinematics_pa_66} we present a plot of the kinematical misalignment for the new sample of 18 galaxies, in addition to the \sauron\ sample. The sample of specials contain four additional slow rotators NGC~4168, NGC~4261, NGC~4365 and NGC~4406, which display a very clear rotation along the apparent minor axis and are certainly not oblate bodies (see \citealt{dav01,sta04} for a detailed discussion about NGC~4365, but see \citealt{vdb07}). These new objects confirm that some of the slow-rotators must be triaxial. A previous indication that large kinematics twists tend to appear only in the boxy slowly rotating galaxies was found by \citet{kor96}. The additional fast-rotators still show no significant kinematical misalignment ($\Psi\la5^\circ$), the only exception being NGC~2679 which has a morphology dominated by the obvious oval shape of a nearly face-on bar. Even the face-on barred galaxy NGC~6548 has a kinematical PA which well agrees with the PA of the large scale stellar disk. In total, out of 49 fast-rotators presented in this paper, the only two that show significant kinematical misalignment, and are thus not consistent with axisymmetry, possess strong bars or show signs of recent interaction. This provides strong evidence that {\em all} the fast-rotators constitute a homogeneous family of flattened and nearly oblate systems (sometimes with embedded bars).

\subsection{Caveats}

A number of caveats and possible improvements apply to the results we discussed in the previous sections:
\begin{description}
\item[\bf Sample size and selection:] The sample of E/S0 galaxies we consider in this paper is still relatively small and is not fully representative of the whole galaxy population. The modest size of the sample constitutes by far the largest limitation of the present study.

\item[\bf Field coverage:] Although the available kinematics extends to larger distances and has higher $S/N$ than what was used in previous studies of the \vse\ diagram, it is still spatially limited and does not sample the whole galaxy, as would be required for a rigorous interpretation of the \vse\ diagram. The comparison between the modelling results, which treat the limited spatial coverage in detail, and the \vse\ results (\refsec{sec:vsigma}), and the test with isotropic models  (Appendix~\ref{sec:test_iso}), all indicate however a general agreement, suggesting the results are reliable. Another effect is that slow-rotators tend to have larger \re\ than the fast-rotators.  For this reason the former are not spatially as well sampled as the latter. We verified that the conclusions would not change if we restricted our analysis to within $\re/2$ for all galaxies.

\item[\bf Bars:] The \sauron\ representative sample includes a number of barred galaxies. Since their complex structure and dynamics \citep[e.g.][]{sel93} can not be represented in detail by oblate spheroids, and axisymmetric models likely provide only a very rough approximation, it is therefore unclear how bars affect correlations derived under the oblate assumption. Most strong bars must however have been rejected by the selection criteria for the $24$ galaxy in our modelling subsample, so they are unlikely to be the primary driver behind the observed trend between $\beta$ and $\epsilon_{\rm intr}$.

\item[\bf Triaxiality:] The slow rotating galaxies are likely weakly triaxial \citep{dez91} and the assumption of axisymmetry in our models certainly introduces a bias in the recovered anisotropy. As it appears that the triaxiality of these systems is not strong, at least in their central regions, the inferred anisotropy is likely to be representative of what we would measure from a full triaxial model \citep{vdb07}. A more detailed investigation of this issue will be presented in a future paper.

\end{description}

\section{Comparison with previous work}

\subsection{\vse\ diagram from long-slit kinematics}

In this section we compare the results on the \vse\ diagram derived from our \sauron\ integral-field stellar kinematics to previous results obtained from long-slit kinematics. The \vse\ diagram from long-slit data of E galaxies was first presented and discussed by \citet{ill77} and \citet{bin78}. It was later extended to galaxy bulges by \citet{ki82}. A widely known form of the diagram is the one presented in \citet{dav83}, which is based on a compilation of long-slit kinematical measurements for 50 elliptical galaxies and bulges. The compilation was extended by \citet{ben92} and many new observations were added by \citet{ben94}. The largest homogeneous set of long-slit \vs\ and ellipticity values is currently constituted by the 94 measurements for elliptical galaxies by R.~Bender (private communication). We will refer to this sample as that of \citet{ben94}, although it was later extended with new unpublished measurements.

To perform a comparison between long-slit and integral-field \vs\ measurements one has to take into account that the two techniques measure two different quantities and cannot be compared in a rigorous way. From integral-field data one measures a luminosity-weighted quantity (equation~[\ref{eq:vsigma}]), which is closely related to the corresponding quantity appearing in the tensor virial equations and can be used directly in a quantitative way. From long-slit kinematics, one usually only measures the quantity $V_{\rm max}/\sigma_0$, where $V_{\rm max}$ is the maximum observed velocity (usually on the galaxy major axis), while $\sigma_0$ is an average value of the velocity dispersion inside $\re/2$. This value cannot be linked to the expression appearing in the tensor virial equations without some rather arbitrary assumptions or `calibration' with models.

From theoretical considerations, \citet{bin05} provides a very rough connection between the two \vs\ determinations. He shows that the long slit $V_{\rm max}/\sigma_0$ measurements agree with the isotropic line, in the old formalism, if the isotropic line (his equation~[9]) is scaled by a factor $\pi/4$ so that for an oblate isotropic galaxy $\vs=0.99$ at $\varepsilon=0.5$. In his new integral-field formalism the isotropic line is a factor $\sqrt{2}$ lower (his equation~[26]) than that in the old one when $\alpha=0$, so that for an oblate isotropic galaxy $\vs=0.89$ at $\varepsilon=0.5$. For our adopted value $\alpha=0.15$ (Appendix~\ref{sec:test_iso}) one has $\vs=0.79$ at $\varepsilon=0.5$ so that the expected combined decrease from long-slit data to integral-field one would be a factor $(\vs)_e\sim0.8\, (V_{\rm max}/\sigma_0)$.
However, as also pointed out by \citet{bin05}, the precise form of this equation depends on the detailed velocity distribution in a galaxy, as well as on the field coverage of the observations, and should not be expected to be quantitatively accurate. This was in fact his motivation to introduce a new formalism for integral-field data.

\begin{figure}
    \plotone{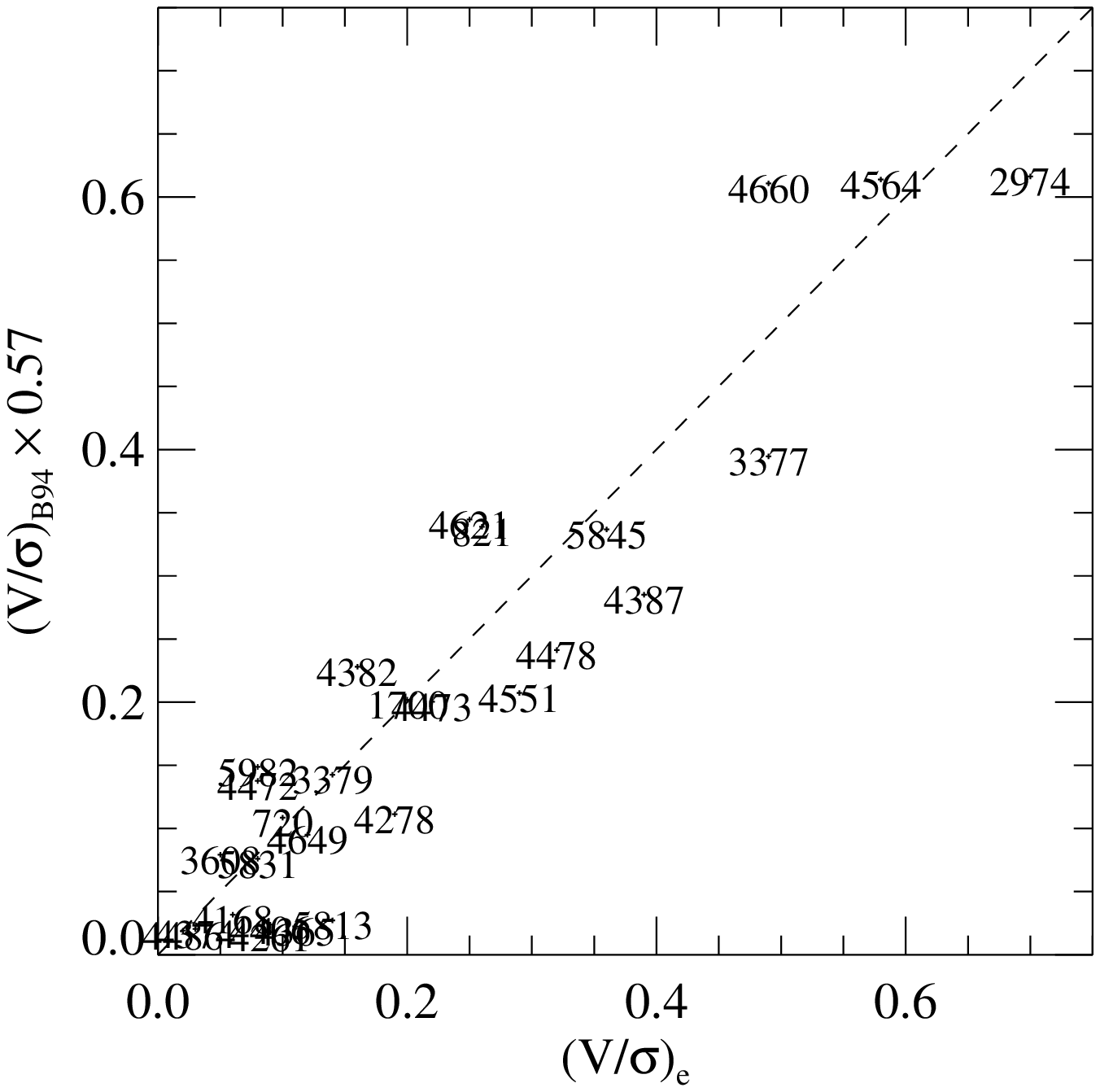}
    \caption{Comparison between the \sauron\ integral-field $(\vs)_{\rm e}$ measurements and the corresponding $V_{\rm max}/\sigma_0$ values by \citet{ben94}, corrected with \refeq{eq:vsigma_scaling} to make them comparable to integral-field observations.}
    \label{fig:comparison_bender94_vsigma}
\end{figure}

\begin{figure}
    \plotone{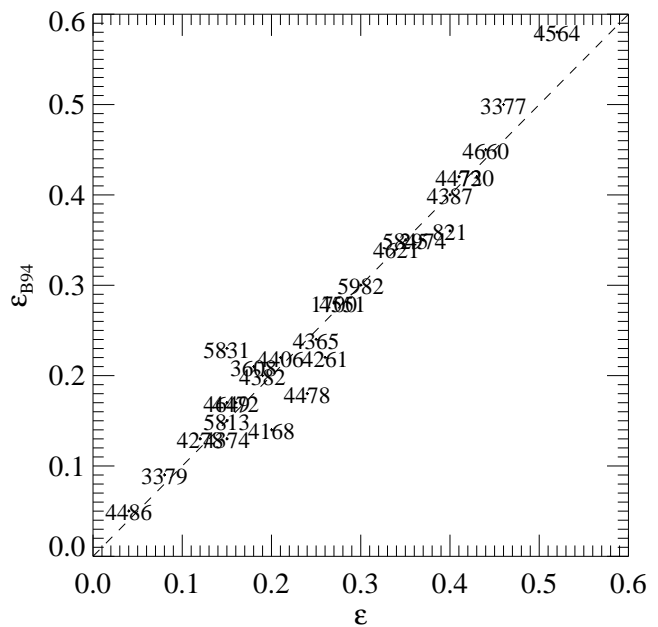}
    \caption{Comparison between the luminosity-weighted values of ellipticity $\varepsilon$ derived for the present paper and the corresponding $\varepsilon_{\rm B94}$ values by \citet{ben94}.}
    \label{fig:comparison_bender94_eps}
\end{figure}

To avoid the approximations and assumptions in the above theoretical derivation, and to be able to compare directly our \sauron\ \vs\ values with those measured with long-slit observations by \citet{ben94}, we selected the 28 galaxies in common between the two samples and we determined the best scaling factor by least-squares fitting (\reffig{fig:comparison_bender94_vsigma}). We found a best-fitting relation
\begin{equation}
\left(\frac{V}{\sigma}\right)_{\rm e} \approx0.57\frac{V_{\rm max}}{\sigma_0},
\label{eq:vsigma_scaling}
\end{equation}
which is $\sim30\%$ lower than the above rough theoretical estimation. Although the differences in individual cases are not insignificant, there appears to be a general agreement between the \sauron\ and (scaled) long-slit \vs\ determinations. The main exception are the galaxies with $\vs\la0.1$, where the long-slit data tend to measure systematically smaller values. A comparison for the ellipticity values measured by us and by \citet{ben94} is shown in \reffig{fig:comparison_bender94_eps}. There is a very good agreement between the ellipticities measured with our different methods.

\begin{figure}
    \plotone{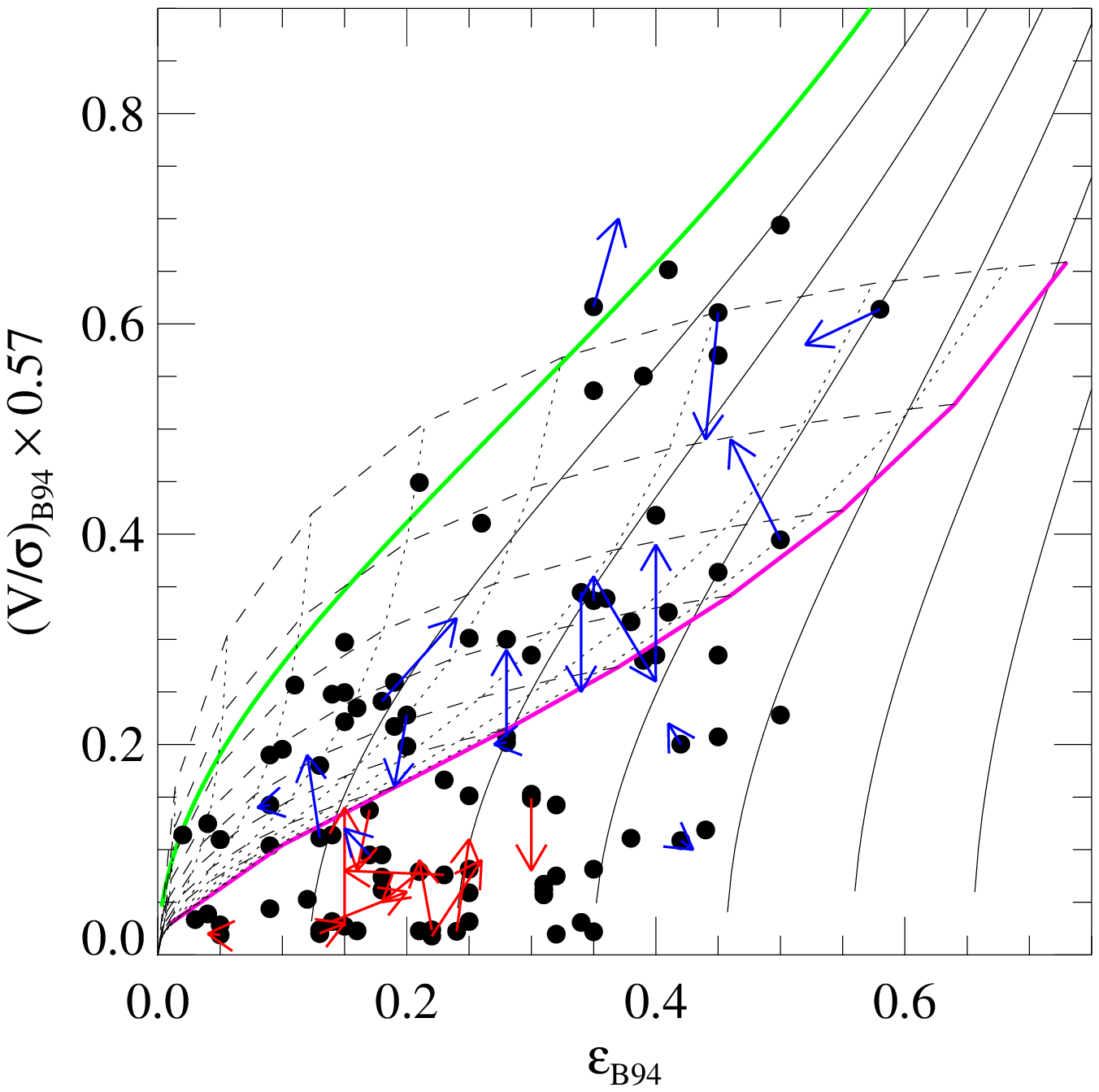}
    \caption{\vse\ diagram from long-slit kinematics. Same as \reffig{fig:v_over_sigma_lines} but for the 94 $V_{\rm max}/\sigma_0$ long-slit measurements and $\varepsilon$ determinations by \citet{ben94}. The $V_{\rm max}/\sigma_0$ values have been corrected with \refeq{eq:vsigma_scaling} to make them comparable to integral-field observations. For the 28 galaxies for which \sauron\ observations also exist, the arrows indicate the location on the diagram of the corresponding value of \reffig{fig:v_over_sigma_66}. The red and blue lines refer to the slow and fast rotators respectively, classified according to the \sauron\ kinematics (Table~\ref{tab2}).}
    \label{fig:comparison_bender94}
\end{figure}

In \reffig{fig:comparison_bender94} we plot the 94 long-slit \vs\ measurements by Bender, rescaled as in \refeq{eq:vsigma_scaling}, and his corresponding ellipticity determinations, on top of the same lines as in \reffig{fig:v_over_sigma_lines}. For the 28 galaxies in the Bender sample for which we have \sauron\ observations, we also show with an arrow the location of the corresponding \sauron\ \vs\ and $\varepsilon$ values. This plot shows that the use of integral-field data does not systematically change the distribution of galaxies on the \vse\ diagram. The differences between the \sauron\ \vse\ diagram and the one in \reffig{fig:comparison_bender94} are discussed in the following section.

The comparison of this section shows that the long-slit observations already provided a reasonable approximation to the more rigorous \vs\ that can be obtained from integral-field kinematics. However integral-field kinematics provides many advantages over long-slit measurements for the study of the \vse\ diagram: (i) The dynamical models can only be uniquely constrained by integral-field kinematics; (ii) \vs\ can be compared to the theoretical value in a rigorous way, without the need for an ad-hoc scaling, allowing for a quantitative comparison with modelling results, in individual cases, as we did in \refsec{sec:vsigma}; (iii) the ability to accurately determine the kinematical misalignment allows us to recognise the different shape distribution of the fast/slow-rotators; (iv) the possibility to rigorously measure the specific angular momentum adds the crucial ability to distinguish between fast and slow-rotators, independently from the \vse\ diagram itself (Paper~IX). A proper kinematical classification of the galaxies is the key for the interpretation of the \vse\ diagram.

\subsection{Sample selection effects on the \vse\ diagram}
\label{sec:selection}

The comparison between the \sauron\ \vse\ diagram of \reffig{fig:v_over_sigma_lines} and the corresponding one by \citet{ben94} in \reffig{fig:comparison_bender94} shows clear differences in the distribution of the observed galaxies. The reason for these differences are easily understood as due to very different selection criteria in the two samples.

The early-type galaxies of the \sauron\ survey were selected to have an absolute blue magnitude $M_B\la -18$, and in this luminosity range to include 24 objects in both the E and S0 morphological classes, to be uniformly sampled in $M_B$ and in the observed ellipticity \citep[as given by][hereafter RC3]{dev91}. By construction, this sample provides a uniform coverage of the \vse\ diagram, but does not give a representative sample of early-type galaxies in the given magnitude range. In fact the true observed shape distribution of E/S0 galaxies displays a clear maximum at an ellipticity $\varepsilon\approx0.25$ and lacks very round and very flat galaxies \citep[e.g.][]{lam92}. The requirement of a uniform sampling in $\varepsilon$ obviously produces an excess of very round and very flat galaxies with respect to the general galaxy population. An additional complication arises from the fact that the \sauron\ selection was done according to the RC3 ellipticities, which are measured at large radii. The ellipticity in galaxies is generally not constant and its variation depends on galaxy luminosity, in the sense that luminous galaxies tend to become rounder in their central parts than in their outer parts \citep{ryd01}. The $\varepsilon$ used in the \vse\ diagram are those measured in the central regions, where the kinematics is available, and not those of RC3.
As a result of this, the \sauron\ selection produces a distribution in $\varepsilon$ which is difficult to quantify.

Although the \sauron\ sample selection, by construction, erases any information given by the $\varepsilon$ distribution, one may naively expect the distribution in \vs\, for any given $\varepsilon$ interval, to be representative of the general E/S0 population, as no explicit selection was performed in \vs. In practice, this is not the case because of the morphological selection, which implicitly introduces a selection on \vs. In particular Table~\ref{tab2} shows that out of the 24 galaxies classified as E from RC3, 14 (58\%) are classified as fast-rotators and 10 (42\%) as slow-rotators. For the S0 galaxies 23 (96\%) are fast-rotators and only 1 (4\%) is a slow-rotator. \reffig{fig:v_over_sigma_lines} shows that the fast-rotators tend to lie above the magenta line, at any given $\varepsilon$, while the slow-rotators are in the bottom-left part of the \vse\ diagram. This implies that the selection of S0 galaxies will guarantee objects above the magenta line in the \vse\ diagram, while E galaxies will provide an equal number of objects above the magenta line and in the bottom-left part of the diagram. The morphological selection thus defines the allowed ranges of \vs\ of any galaxy. The requirement of an equal number of E and S0 galaxies in every interval of $\varepsilon$ specifies a certain distribution of galaxies in the \vse\ diagram, which is not representative of the early-type galaxies population. These selection biases are very difficult to estimate and to correct for from the present sample, as they depend on the unknown distribution of fast/slow-rotators as a function of intrinsic galaxy shape and morphological classifications.

The morphological selection also affects the appearance of the \vse\ diagram as obtained from the long-slit measurements by \citet{ben94}. That sample is in fact constituted only by E galaxies. As originally pointed out by \citet{dav83}, clearly stated by \citet{kor96} and strongly confirmed in this paper and Paper~IX, the E class constitutes a quite heterogeneous group, made by less than half of genuine non-rotating ellipticals (the slow-rotators) and the rest by disk-like systems, resembling S0 galaxies (the fast-rotators). The appearance of the \vse\ diagram of E galaxies thus depends on the fraction of fast-rotators that are (mis)classified as E in different $\varepsilon$ intervals. A direct comparison with the \sauron\ \vse\ diagram for the E galaxies alone (open circles) can be seen in \reffig{fig:vsigma_morphology}. Although the \sauron\ sample has only 25\% of the number of E of \citet{ben94}, the distribution of the galaxies appears qualitatively similar.

The \vse\ values for the E galaxies of \citet{ben94} in \reffig{fig:comparison_bender94} span the same region as those of \sauron\ of \reffig{fig:v_over_sigma_lines} or \ref{fig:v_over_sigma_66}. However there are three main qualitative differences between this diagram and the \sauron\ one, which can be understood as due to differences in the sample selection: (i) the E selection in the long-slit sample poses an upper limit on $\varepsilon\la0.5$, as otherwise the galaxies would be classified as S0. (ii) The lack of S0 in the long-slit sample implies a smaller fraction of fast-rotating galaxies (above the magenta line), compared to the \sauron\ \vse\ diagram. (iii) The fact that there was no explicit selection on $\varepsilon$ in the long-slit sample gives a smaller fraction of very round galaxies, compared to the \sauron\ sample. Moreover older samples of \vs\ values tend to prefer flatter E, as these were thought to be most likely to be seen edge-on, and consequently better consistent with the assumption of the \vse\ diagram.

It is important to obtain an unbiased view of the \vse\ diagram, from a larger statistical sample, to be able to compare the observed distribution of fast-slow rotators, with the predictions from different galaxy formation scenarios. For this, one needs a sample which is only selected in absolute magnitude and is not biased by the morphological selection, as the currently available samples. Some selection in morphological type is still required, as the \vse\ diagram looses its meaning for spiral galaxies, which have a clear bulge and a thin disk component with very different ellipticity. However, contrary to the problematic classification into E or S0 galaxies, the separation between early-types galaxies and spirals is robust and clearly bimodal \citep{str01,bal04}. Only with this larger and unbiased sample of E/S0 galaxies will we be able to understand whether the fast-rotators all really fall within the envelope indicated by the current study, and whether the distribution of the slow/fast-rotators shows a bimodality in the \vse\ diagram, as the current data seem to suggest.

\subsection{Spurious increase in anisotropy of slow-rotators?}

In \refsec{sec:selection} we showed that our \sauron\ data on the \vse\ diagram are generally consistent with previous long-slit works. The current study however seems inconsistent with previous studies \citep[e.g.][]{dav83,bin87,kor96} in its finding that the massive and slowly rotating galaxies are {\em not} more anisotropic than the generally fainter and fast-rotating galaxies (e.g.\ \reffig{fig:sigma_beta}).

Part of the difference between the current work and the previous ones may lie in the sample selection, which here includes S0 galaxies, as they cannot be distinguished from the other fast-rotators. The rest can be understood as due to (i) the use of the parameter $(\vs)^\star$ as a measure of galaxy anisotropy and (ii) the neglect of inclination effects on the \vse\ diagram.

\begin{figure}
    \plotone{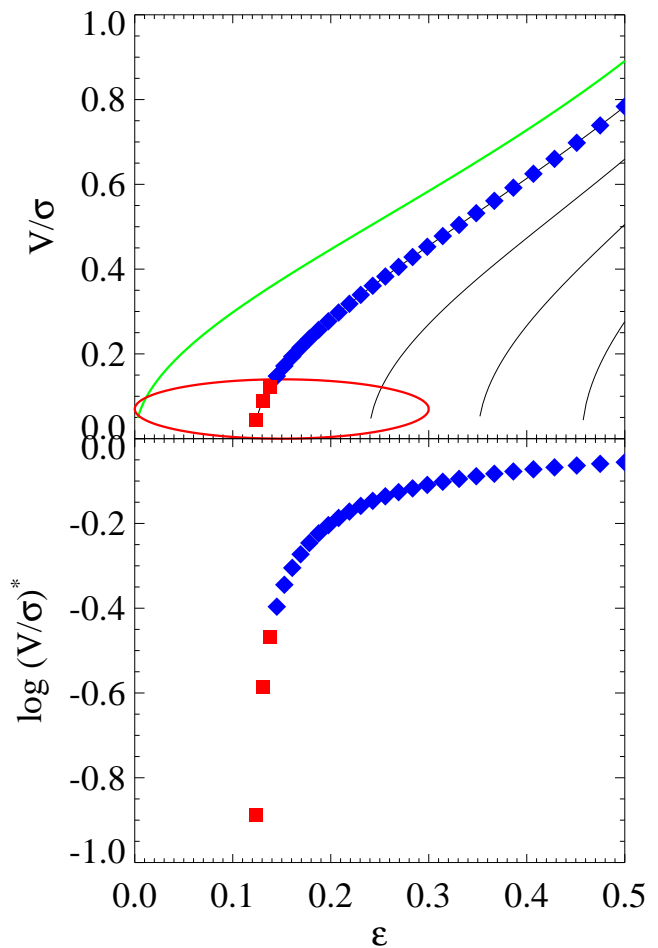}
    \caption{{\em Top Panel:} \vse\ diagram. The meaning of the solid lines is as in \reffig{fig:v_over_sigma_lines}. The blue diamonds and red squares represent the location of edge-on galaxies, with different intrinsic ellipticity $\varepsilon_{\rm int}$ but the same anisotropy $\delta=0.1$. The red ellipse indicates the region where the slow-rotators are generally observed (\reffig{fig:v_over_sigma_lines} and \ref{fig:comparison_bender94}). {\em Bottom Panel:} Location of the galaxies of the top panel in the $(\log(\vs)^\star,\varepsilon)$ diagram. The diamonds and squares represent galaxies with the same small anisotropy, so one should expect $\log(\vs)^\star\approx0$. There is however a dramatic decrease of $\log(\vs)^\star$ for the objects which fall close to the region of the slow-rotators (red squares), which is not associated to any variation in the anisotropy.}
    \label{fig:v_over_sigma_star}
\end{figure}

The anisotropy parameter was defined by \citet{kor82b,dav83,ben90} as
\begin{equation}
(\vs)^\star\equiv\frac{\vs}{(\vs)_{\rm iso}},
\end{equation}
i.e.\ the ratio between the observed \vs\ value for a galaxy and the predicted value $(\vs)_{\rm iso}$ for an oblate galaxy with isotropic velocity dispersion tensor \citep{bin78}. As pointed out by \citet{bin05}, $(\vs)^\star$ is not a reliable way of measuring anisotropy in a galaxy and should be replaced by the real anisotropy value $\delta$. The problems of $(\vs)^\star$ as a measure of anisotropy $\delta$ are the following: (i) galaxies with the same $\delta$ but with different intrinsic ellipticity $\varepsilon_{\rm int}$, can appear to have dramatically different $(\vs)^\star$ when seen edge-on, and even more different $\log(\vs)^\star$. In particular the rounder galaxies will have a systematically lower $(\vs)^\star$ at any given $\delta$ (\reffig{fig:v_over_sigma_star}). (ii) A galaxy with a given $\delta$ and $\varepsilon_{\rm int}$, will appear to have a lower $(\vs)^\star$ at lower inclinations ($i=90^\circ$ being edge-on) as shown e.g.\ in \reffig{fig:v_over_sigma_lines} \citep[see also][]{bur05}.

The slow-rotators generally have observed ellipticities $\varepsilon\la0.3$ (\reffig{fig:v_over_sigma_lines} and \ref{fig:comparison_bender94}), which implies they are intrinsically not very flat. The fast rotators can be intrinsically flat. Given this difference in the shape of the fast and slow-rotators, both the ellipticity and inclination effects mentioned in the previous paragraph act to systematically decrease the observed $(\vs)^\star$ of the slow-rotators with respect to the fast-rotators, giving the misleading impression of a strong increase of the anisotropy. We conclude that $(\vs)^\star$ should not be used  as a measure of anisotropy.

\section{Discussion}
\label{sec:discussion}

\subsection{Early-type galaxies in the nearby universe}

Observations at high redshift, combined with detailed measurements of the cosmic microwave background, have provided a scenario for galaxy formation in which the Universe is dominated by dark matter of unknown nature. The general picture for the assembly of the dark matter seems to be rather well understood and is described by a hierarchical process driven by the sole influence of gravity  \citep[e.g.][]{spr05}. Much however still has to be learned about the way the luminous matter, namely the stars and galaxies that we actually observe, form from the accretion of gas within the dark matter potential. The complication comes from the fact that these processes are driven by the hydrodynamic and chemistry of the gas, combined with complex feedback processes of the radiation on it.

One way to constrain the formation of the luminous matter in galaxies is by trying to uncover the `fossil record' of that process in nearby galaxies, which can be observed in great detail due to their vicinity. A key element to differentiate between different galaxy-formation model is the degree of anisotropy and the amount of rotation in early-type galaxies, in relation to other galaxy properties like luminosity, surface brightness profiles or chemical composition \citep[e.g.][]{bin78,dav83}.
Observations show that early-type galaxies can be broadly separated into the most massive giant ellipticals on one side, which are red, metal-rich, slowly-rotating, and have shallow central photometric profiles, and the fainter objects on the other side, which tend to be bluer, metal-poor, are dominated by rotation, have cuspy profiles and may contain disks \citep{dav83,jaf94,kor96,fab97,lau05}

A first factor determining these observed differences in the structure of early-type galaxies appears to be the amount of dissipation during the last galaxy merger. In particular more anisotropic and slowly rotating galaxies result from predominantly collisionless major mergers, while faster rotating galaxies are produced by more gas-rich mergers, where dissipation plays an important role \citep{kor89,ben92,fab97}. The prevalence of gas-poor mergers in the hierarchical formation of massive galaxies \citep{kho05} would help explain the fact that slowly rotating galaxies are more common in the high-mass galaxy range. A second factor driving the galaxy structure seems to be the mass fraction of the merger components. Equal-mass galaxy mergers tend to produce more slowly rotating galaxies than mergers of significantly different mass ratio \citep{naa99,naa03}. A third factor appears to be the feedback on the gas, during a merger event, produced by the radiation emitted by a central AGN or by supernovae winds \citep[e.g.][]{gra04,spr05a}.

A complete understanding of the formation of early-type galaxies, and the relative role of the different factors, requires the treatment of the above phenomena in a cosmological context, within the hierarchical merging process. This still has to be attempted in a rigorous way as current computer simulations cannot resolve the many orders of magnitude scales that are involved in these processes. This implies that instead of describing the gas via accurate hydrodynamical and radiative transfer equations, the models use approximate and very simple relations, based on empirical relations, to describe the process. This paper and the companion Paper~IX provide new observational constraints for the upcoming models. In the next section we speculate on the formation scenarios for the observed characteristics of early-type galaxies that we derived.

\subsection{Nature of the fast and slow rotators}

The fast rotators seem consistent with galaxies with a significant disk component, which experienced minor mergers and accreted a significant amount of gas \citep{naa99,naa03,bou05,cox06,jes06,nov06}. The accreted gas, which settles in the galaxy equatorial plane and forms stars, would explain the metallicity enhancement in the plane of the disk (Paper~VI) and the nearly perfect alignment of the angular momentum of the different, sometimes counterrotating, stellar structures (\refsec{sec:understand}). The minor mergers may explain the fact that rounder galaxies have a larger $\sigma_z$  dispersion orthogonal to the disk plane (along the $z$-axis). The disk instabilities triggered by minor mergers can in fact deviate stellar orbits out of the disk plane and produce an increase in the $\sigma_z$ dispersion and a corresponding thickening of the disk. The trend may also be related to the two-components nature of the fast-rotators. Disks may be more anisotropic than bulges and are more prominent in flat galaxies (\refsec{sec:bulges}). Numerical simulations seem able to produce the observed level of anisotropy in these systems \citep{bur05}, but it is unclear for what conditions they will reproduce the observed trend between anisotropy and flattening.

Although we do not find any physical distinction between E fast-rotators with disky isophotes and S0 fast-rotators with an obvious disk component, and this suggests a common formation scenario for the whole class of objects, the current observations do not necessarily require a common origin.

Originally the slow rotators were thought to form at high redshift via equal mass mergers of disk galaxies \citep{bar88}. The N-body simulations were able to produce some slowly rotating systems that appeared like ellipticals, however it became clear that this simple picture could not explain the observed appearance of slowly rotating ellipticals as a class \citep{her92,naa03}. In addition, current numerical simulations of gas-free major mergers seem to generate large fractions of strongly triaxial or prolate galaxies ($0.2\la\varepsilon\la0.6$) than we do {\em not} observe \citep[e.g.][]{boy05,bur05,gon05,cox06}. A missing ingredient in these models is the cosmological context, which produces sequences of mergers. Also lacking is a detailed treatment of the gaseous component, which is likely to play an important role in the early universe. Major gas-rich mergers, regulated by the feedback of a powerful central AGN, have been suggested to explain the scaling relations of early-type galaxies \citep{rob06} and to reproduce the dichotomy \citep[e.g.][]{fab06,cat06} in the colour distribution of galaxies as a function of their luminosity \citep{str01,bal04}.

It is unclear why the slow rotators in our sample possess the low degree of anisotropy that we observe in their central regions. A process that could in principle explain both the relatively small anisotropy and the roundish shape of the slow rotators would be the global transition to chaos, as can be produced by central supermassive black holes \citep{mer96}. Current estimates of black holes mass fractions \citep{har04} seem however too small to produce significant effects. The transition to chaos and the symmetrisation of the major triaxial merger remnant could be significantly enhanced by the gas inflow and its nuclear concentration during the initial phase of the mergers, before the gas is expelled from the galaxies by the AGN activity \citep{bar88}. Another process that could explain the isotropy and shape of the slow-rotators, especially for the most massive galaxies found near the centre of cluster potentials, is a sequence of radial mergers \citep{coo05}.

\subsection{Connection with higher redshift observations}

The detailed observations of the structure of the small sample of nearby galaxies that we discuss can be related to much less detailed observations at higher redshift obtained with much larger survey. From the analysis of a large sample of galaxies in the nearby universe, observed by the Sloan Digital Sky Survey, it was found that galaxies are well distinct in terms of their colour, in the sense that they can be clearly separated in a so called `blue cloud' and a `red sequence' \citep{str01,bal04}. This discovery, and the subsequent realisation that the bimodality can be traced back in time at higher and higher redshift \citep{bel04}, allowed a dramatic improvement in the detailed testing of galaxy formation scenarios. It is now believed that the gas-free and dead, red-sequence galaxies form by merging of gas-rich, star-forming, blue-cloud galaxies, followed by a rapid ejection of the gas, due to the feedback from a central supermassive black hole \citep[e.g.][]{fab06}.
In a merger event between blue gas-rich galaxies the gas tends to settle onto a plane and form a disk, so that the end result of the merger, after the gas has been expelled from the system, will be a red, purely stellar system dominated by rotation. But mergers can still happen between red gas-poor galaxies, in which case simulations show that the resulting red galaxy will show little or no rotation.

We speculate that the two classes of fast and slow-rotating galaxies that we described may be the relics of the different formation paths followed by early-type galaxies when loosing their gas content and moving from the `blue cloud' to the `red sequence', in the `quenching' scenario of \citet{fab06}. In that picture the fast-rotators are naturally associated with the spiral galaxies whose disks were quenched by ram-pressure stripping or other ram-starvation processes, and subsequently experienced mainly minor mergers. The slow-rotators can be generated by a sequence of mergers involving little amount of gas along the red sequence, or due to major violent gas rich mergers in which the gas component was rapidly expelled by a starburst or a central AGN.

\section{Conclusions}

We analysed the orbital distribution of elliptical (E) and lenticular (S0) galaxies using observations of the stellar kinematics within one effective (half light) radius obtained with the \sauron\ integral-field spectrograph. We constructed the \vse\ diagram for the 48 E/S0 of the \sauron\ sample. For a subsample of 24 of the galaxies, consistent with axisymmetry. we use three-integral axisymmetric Schwarzschild models to recover the detailed orbital distribution.

We used a new classification of early-type galaxies based on the stellar kinematics (Paper~IX), to divide the sample into two classes of objects with and without a significant amount of angular momentum per unit stellar mass. We refer to these two types of object as ``slow-rotators'' and ``fast-rotators'' respectively. The slow-rotators  are more common among the brightest systems, and are generally classified as E from photometry alone. The fast-rotators are generally fainter and are classified either E or S0.

From the combined analysis of the dynamical models and the \vse\ diagram we find that, in the central regions (within $\sim1\re$) of the galaxies which we sample with our kinematics:
\begin{enumerate}
\item The slow rotators can display misalignment between the photometric and
kinematic axes, indicating that as a class they are triaxial. Those in
our sample tend to be fairly round ($\varepsilon\la0.3$), and span a range of moderate anisotropies (Binney 1978 anisotropy parameter $\delta\la0.3$).

\item The fast-rotators can appear flattened ($\varepsilon\la0.7$), do not show significant kinematical misalignment (unless barred or interacting), which indicates they are all nearly axisymmetric, and span a large range of anisotropies ($\delta\la0.5$). {\em The main physical difference between fast and slow-rotators is not their degree of anisotropy, but their intrinsic shape.}

\item The measured anisotropy appears to be due to a flattening of the velocity ellipsoid in the meridional plane ($\sigma_R>\sigma_z$), which we quantify with the $\beta$ anisotropy parameter. While the velocity dispersion tends to be similar along any direction orthogonal to the symmetry axis ($\sigma_R\sim\sigma_\phi$).

\item We find a general trend between the anisotropy in the meridional plane $\beta$ and the galaxy intrinsic ellipticity $\varepsilon_{\rm intr}$. Flatter galaxies tend to be more anisotropic.

\item The results from the models are consistent with the values of the anisotropy inferred from the \vse\ diagram of the whole \sauron\ sample of 48 E/S0, which we determined from our integral field kinematics. We find that the distribution of the fast rotating galaxies is well described as a family of nearly oblate galaxies following a trend between flattening and anisotropy. This result is confirmed when we include in the sample 18 additional E/S0 galaxies observed with \sauron.
\end{enumerate}

Although the sample of early-type galaxies considered in this work constitutes the first significant sample for which the \vse\ diagram and the anisotropy could be investigated with integral-field stellar kinematical data, the main limitation of our work is the relatively small number of galaxies and the complicated sample selection effects. Even so, as we discussed in \refsec{sec:discussion}, we are able to reach some provisional conclusion on the formation of early-type galaxies. We reviewed the usefulness of the \vse\ diagram to test galaxy-formation models. A similar analysis, performed on a much larger sample of galaxies could provide much stronger constraints on the process by which galaxies form.

\section*{Acknowledgements}

We thank Andi Burkert and Marijn Franx for very useful comments, Ralf Bender for providing his latest long-slit \vse\ measurements and Ortwin Gerhard for providing his numerical values of the anisotropy. We thank the referee, John Kormendy, for a detailed report which helped significantly improve the presentation of our work.
The \sauron\ project is made possible through grants 614.13.003, 781.74.203, 614.000.301 and 614.031.015 from NWO and financial contributions from the Institut National des Sciences de l'Univers, the Universit\'e Claude Bernard Lyon~I, the Universities of Durham, Leiden, and Oxford, the British Council, PPARC grant `Extragalactic Astronomy \& Cosmology at Durham 1998--2002', and the Netherlands Research School for Astronomy NOVA. MC acknowledges support from a VENI award (639.041.203) by the Netherlands Organization for Scientific Research (NWO) and a PPARC Advanced Fellowship (PP/D005574/1). RLD is grateful for the award of a PPARC Senior Fellowship (PPA/Y/S/1999/00854) and postdoctoral support through PPARC grant PPA/G/S/2000/00729. The PPARC Visitors grant (PPA/V/S/2002/00553) to Oxford also supported this work. JFB acknowledges support from the Euro3D Research Training Network, funded by the EC under contract HPRN-CT-2002-00305. GvdV acknowledges support provided by NASA through grant NNG04GL47G and through Hubble Fellowship grant HST-HF-01202.01-A awarded by the Space Telescope Science Institute, which is operated by the Association of Universities for Research in Astronomy, Inc., for NASA, under contract NAS 5-26555. Photometric data were obtained (in part) using the 1.3m McGraw-Hill Telescope of the MDM Observatory. The \sauron\ observations were obtained at the William Herschel Telescope, operated by the Isaac Newton Group in the Spanish Observatorio del Roque de los Muchachos of the Instituto de Astrof\'{\i}sica de Canarias. This research has made use of the NASA/IPAC Extragalactic Database (NED). We acknowledge the usage of the HyperLeda database (http://leda.univ-lyon1.fr).

{}

\appendix

\section{Testing anisotropy with two-integral models}
\label{sec:jeans}

The three-integral axisymmetric models of \refsec{sec:momemnts} show that, in the central regions that we constrain with our integral-field kinematics, (i) the flatter fast-rotators are anisotropic, in the sense that their velocity ellipsoid is flatter along the direction of the symmetry axis than along the radial direction $\sigma_z<\sigma_R$; (ii) the rounder slow-rotators are nearly isotropic. In this section we do a qualitative test of the consistency and robustness of these statements with the results of two-integral Jeans models. This test uses a very different and simpler modeling method and a different kinematics extraction. It allows us to verify the robustness of our results against implementation and observational details.

The Jeans models we use assume axisymmetry, constant $M/L$ ratio, and a stellar distribution function DF of the form $f=f(E,L_z)$, where $E$ is the energy and $L_z$ is the angular momentum along the symmetry axis $z$. From the same MGE parameterization of the surface brightness used in the Schwarzschild models (\refsec{sec:models}) and an estimate of the central black-hole mass, we compute the unique predictions for the velocity second moments $\mu_2$, at the best-fitting inclination, using the formalism of \citet{ems94}. See Paper~IV for more details.
The second velocity moments were extracted from the \sauron\ kinematics as $\mu_2\equiv\sqrt{V^2+\sigma^2}$, where $V$ and $\sigma$ are the mean velocity and the velocity dispersion used also in the \vse\ diagram (\refsec{sec:sample}).

\subsection{Anisotropy of the flat fast-rotators}
\label{sec:jeans_fast}

\begin{figure}
    \plotone{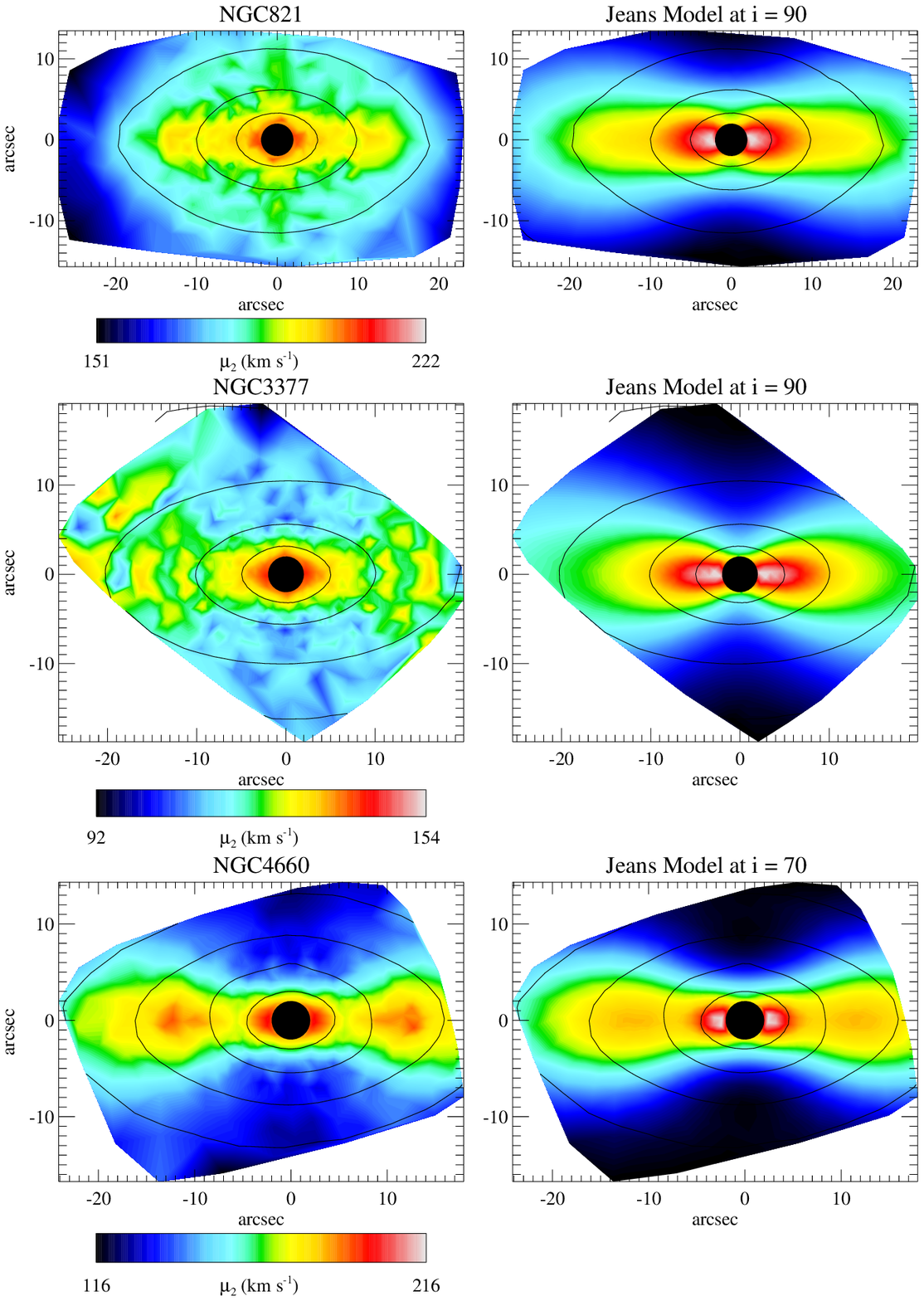}
    \caption{Jeans models of flattened galaxies. Left panels: bi-symmetrised and linearly interpolated maps of the observed second velocity moment $\mu_2\equiv\sqrt{V^2+\sigma^2}$, measured from the \sauron\ stellar kinematics of the galaxies NGC~821, NGC~3377 and NGC~4660. The actual Voronoi bins in which the kinematics was measured are shown in Paper~III. All three galaxies are quite fattened and are constrained by the observed photometry to be close to edge-on. Right panels: two-integral Jeans modelling of the unique projected second moments associated to the MGE mass density distribution, at the best fitting inclination from Paper~IV. In all three cases the models show a larger difference between the $\mu_2$ measured along the major and minor axes than is observed.
    \label{fig:jeans_maps}}
\end{figure}

To test the anisotropy of the flat fast rotators we consider the three galaxies NGC~821, NGC~3377 and NGC~4660, which are constrained by the observed photometry to be close to edge-on. This choice reduces the non-uniquness in the mass deprojection \citep{ryb87} and correspondingly the possible degeneracies in the deprojection of the higher velocity moments. These three galaxies have a $\sigma$ field well resolved by the \sauron\ instrumental resolution. For edge-on galaxies one does not have any direct measurement of the component of the velocity dispersion $\sigma_z$ parallel to the symmetry axis, however the flattening of the velocity ellipsoid is tightly constrained if the galaxy is assumed to be axisymmetric and in equilibrium (same as in the \vse\ diagram). Nearly edge-on galaxies are thus ideal for our test.

The comparison with the Jeans model predictions is shown in \reffig{fig:jeans_maps}. The models provide a reasonable first-order description of the observations, however significant systematic deviations exists. This indicates that, assuming the mass traces the light, the modelled galaxies do not have a DF of the form $f(E,L_z)$. In particular a common characteristics of models is that for all three galaxies they predict too much motion (high $\mu_2$ values) along the galaxies projected major axis with respect to the minor axis. Any two-integral $f(E,L_z)$ model has a velocity-ellipsoid which appears circular in cross section with the meridional plane ($\sigma_R=\sigma_z$ and $\langle v_z v_R \rangle=0$). The observed difference between models and observations indicates that the real galaxies have a larger velocity dispersion than the models, along the $R$ direction, orthogonally to the $z$ symmetry axis. This implies that the velocity ellipsoid must be on average elongated in the $R$ direction. In this way, for nearly edge-on galaxies, the observer sees a larger component of the dispersion along the minor axis, when the radial direction in the galaxy is parallel to the line-of-sight, than along the major axis. A similar finding was interpreted by \citet{van91} as evidence of radial anisotropy and the same reasoning applies in our case.

\subsection{Near isotropy of the round slow-rotators}

\begin{figure}
	\centering
    \includegraphics[width=0.7\columnwidth]{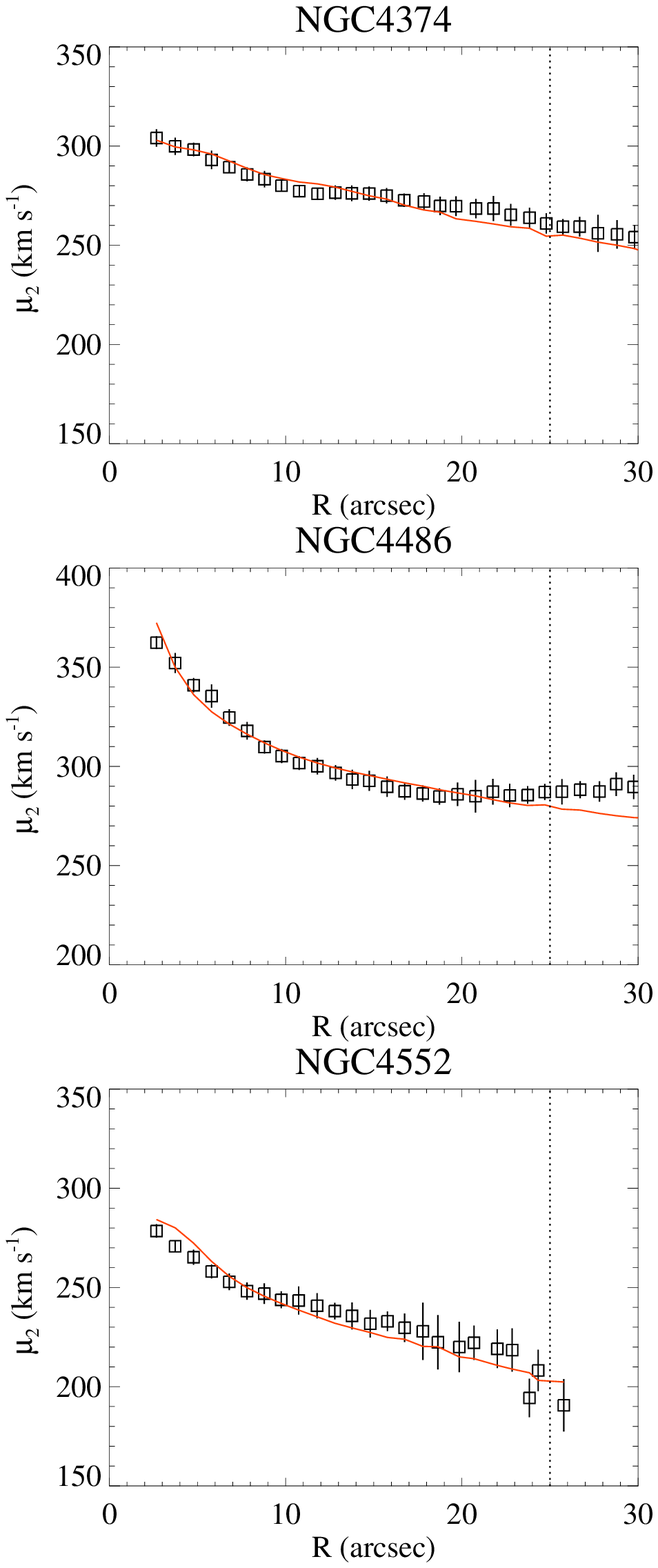}
    \caption{Jeans models of nearly round galaxies. The open squares with error bars represent the observed second velocity moment $\mu_2\equiv\sqrt{V^2+\sigma^2}$, measured from the \sauron\ stellar kinematics of the galaxies NGC~4374, NGC~4486 and NGC~4552. The values and errors have been computed as the biweight mean and standard deviation of the kinematics extracted from the Voronoi bins within circular annuli. The solid line shows the second velocity moment computed from two-integral Jeans models. The dashed vertical line shows the region where the \sauron\ field provides a full spatial coverage of the galaxy. This is the region visualized in \reffig{fig:rotation_all_sample} and from which the mean anisotropy was estimated. The two-integral models provide a very good description of the data within the central region. Given the near spherical symmetry the models are almost fully isotropic in the central regions.\label{fig:jeans_profiles}}
\end{figure}

To test the near isotropy inferred using three-integral models for the rounder slow-rotators of the modelling subsample, we selected the three apparently quite round galaxies NGC~4374, NGC~4486 and NGC~4552. These three galaxies have none or very little evidence for rotation and the velocity dispersion field presents a nearly circular symmetric distribution (Paper~III). We constructed axisymmetric $f(E,L_z)$ Jeans models as in the previous section, taking into account the apparent flattening of the galaxies in the outer regions, however, given the near-spherical shape in the central regions, the models are almost fully isotropic with $f=f(E)$. It is sufficient to compare data and models along circular rings\footnote{The two-dimensional data versus model comparison for NGC~4552 was presented in Fig.~A1 of Paper~IV.}. The \sauron\ data were averaged by computing a biweight mean and standard deviation \citep{hoa83} of the values extracted from the Voronoi bins within radial ranges. The use of integral-field kinematics provides radial profiles of $\mu_2$ with very small scatter.

If the galaxies were nearly isotropic in their central regions, as the Schwarzschild models show, the simple Jeans models, with only $M/L$ as free parameter, should provide a good representation of the observed $\mu_2$ profiles. The comparison between the $\mu_2$ profiles of the data and the isotropic models (\reffig{fig:jeans_profiles}) shows that indeed the models generally reproduce the data within the small uncertainties, and confirms the consistency of the observed galaxies with a nearly isotropic orbital distribution. The galaxies are of course not perfectly isotropic: deviations between data and model do exist, particularly in the outer regions of NGC~4486, and these small deviations from isotropy are consistent with the location of the galaxies below the isotropy line in the \vse\ diagram (\reffig{fig:v_over_sigma_lines}).

\section{Testing the \vse\ diagram with isotropic models}
\label{sec:test_iso}

For a quantitative use of the \vse\ diagram three questions have to be answered: (i) can one measure reliable values of the anisotropy using the \vse\ diagram from integral-field kinematics with limited spatial coverage? (ii) Can one derive realistic anisotropies for models with an ellipticity that varies as a function of radius? (iii) what is the value for the parameter $\alpha$ (equation~[\ref{eq:delta_vs}]) appropriate for realistic galaxy models?

We tried to address some of these questions in \refsec{sec:vsigma}, by comparing the anisotropy derived from the Schwarzschild models with the one measured with the \vse\ diagram. The good agreement between the two determinations showed that the answer to the first two questions is positive. These tests allowed us to verify the reliability of the \vse\ diagram for a wide range of anisotropies. However the approximate nature of the orbit-superposition models does not allow the test to be carried out very accurately.

An alternative would be to use analytic models. The limitation of these tests is that they are easy to carry out only in the isotropic case. This was done by \citet{bin05} who also concluded with a positive answer to the first two questions above. In addition he measured a value of $\alpha\la0.2$ for his test cases. However his adopted models were not entirely realistic, so the results cannot be necessarily generalised to real galaxies.

In this section we revisit what was done by \citet{bin05}, using the MGE models to approximate in detail the surface brightness of actual galaxies. For this test we selected five galaxies which represent some of the different types found in our \sauron\ sample and for which we derived a nearly edge-on inclination in Paper~IV. The galaxy NGC~4486 is a giant elliptical with the largest $\sigma_e$ in our sample. It shows no sign of rotation and has a nuclear cuspy core in the surface brightness profile. NGC~4552 and NGC~5831 are also classified as slow rotators and have central cuspy cores. NGC~3377 and NGC~4621 are classified as fast rotators, are coreless and have disky isophotes.

We computed the projected $V$ and $\sigma$ fields for the five galaxies under the assumption of axisymmetry and self-consistency (i.e. constant $M/L$). In this case, when the distribution function is assumed to be a function $f(E,L_z)$ of the two classical integrals of motion, the energy $E$ and the angular momentum $L_z$ with respect to the symmetry axis, the non-centred second moment $\mu_2$ is uniquely defined by the density, once the inclination $i$ and the $M/L$ are chosen \citep{lyn62,hun77}. If in addition one assumes isotropy of the velocity dispersion tensor, then the first moment $\mu_1$ of the velocity is also uniquely specified. Both moments $\mu_1$ and $\mu_2$ can be computed by solving the Jeans and Poisson equations \citep[e.g.][]{sat80,bin90}. In the case the density is described by the MGE parameterization the moment $\mu_1$, projected on the sky plane, can be evaluated as a double integral using equation~(59) of \citet{ems94}. The projected moment $\mu_2$ is evaluated as a single integral via equation~(61) of the same paper (corrected for a typo as in Paper~IV). We made the standard approximate identification $V\equiv\mu_1$ and $\sigma^2\equiv\mu_2^2-\mu_1^2$.

To be able to measure the parameter $\alpha$ the MGE models need to have constant ellipticity as a function of radius, as the adopted form of the \vse\ diagram precisely applies only in that case. For this we took the MGE parameters as given in Paper~IV but we set all the ellipticities $\varepsilon$ equal to a constant value $\varepsilon=0.2,0.4,\ldots,0.8$. In this way the new MGE models have the same realistic major axis profile of the original galaxy, but have constant $\varepsilon$. To evaluate the Jeans equations in a computationally efficient and accurate way up to large radii, and even for very flat models, we computed them on a grid linear in the logarithm of the elliptical radius and in the eccentric anomaly. This was done by defining a logarithmically-spaced radial grid $r_j$ and then computing the moments at the coordinate positions $(x_{j,k},y_{j,k})=[r_j\cos\theta_k, r_j (1-\varepsilon) \sin\theta_k]$, for linearly spaced $\theta_k$ values. The model was then re interpolated onto a fine Cartesian grid for the estimation of the \vs\ as done on the real integral-field data.

\begin{figure}
\centering
\includegraphics[width=0.9\columnwidth]{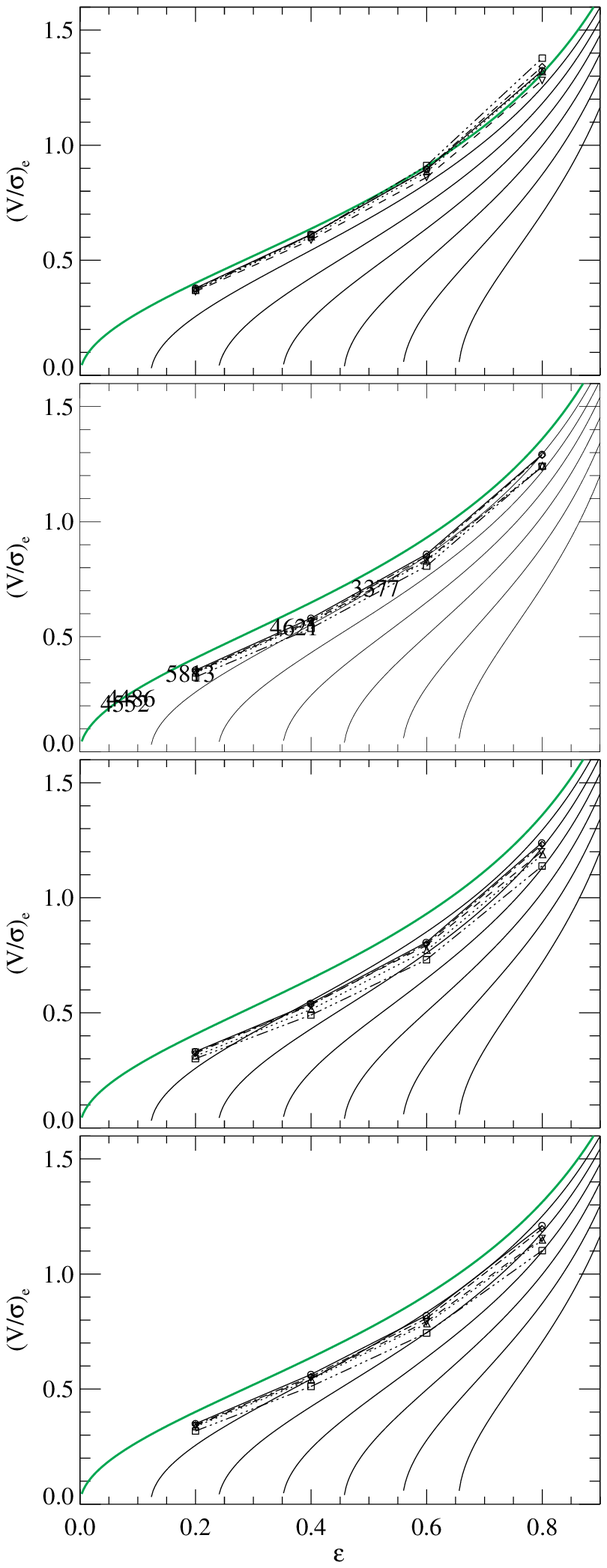}
    \caption{\vs\ of isotropic galaxy models. {\em Top plot:} the \vs\ was measured from the model by integrating it within an ellipse of ellipticity $\varepsilon$ and mean radius $R_\star\equiv a\sqrt{1-\varepsilon}=3\re$, where $a$ is the semi-major axis of the ellipse. {\em Second panel:} the \vs\ was measured within an ellipse of mean radius $R_\star=\re$. The NGC numbers refer to the \vse\ values measured on the isotropic models with variable ellipticity. {\em Third panel:} \vs\ within an ellipse of mean radius $R_\star=\re/2$. {\em Bottom panel:} the \vs\ was measured within circles of radius \re. This plot shows that \vs\ is better recovered along ellipses than along circles of the same area. Moreover the \vs\ values appear to be well recovered up to an ellipse of mean radius \re.}
    \label{fig:isotropic_test}
\end{figure}

In \reffig{fig:isotropic_test} we show the location of the five isotropic galaxy models on the \vse\ diagram. In the top plot the \vs\ was measured from the model by integrating it within an ellipse of ellipticity $\varepsilon$ having mean radius $R_\star\equiv a\sqrt{1-\varepsilon}=3R_{\it e}$, where $a$ is the semi-major axis of the ellipse. For each different $\varepsilon$ the five models have similar \vs\ value and they lie along the theoretical isotropic line if one adopts $\alpha\approx0.20\pm0.03$. In the second panel the \vs\ was measured within an ellipse of mean radius $R_\star=R_{\it e}$, which closely resemble what we are doing on the real \sauron\ data. In this case the models lie slightly below the isotropic line, but the inferred anisotropy is still negligible $\delta\la0.1$. This shows that an ellipse of mean radius $R_{\it e}$ is still sufficient for a reliable recovery of the \vs\ value of realistic galaxy models. This is not the case when the \vs\ is measured within an ellipse of mean radius $R_\star=R_{\it e}/2$, as in this case the values start being significantly underestimated ($\delta\ga0.2$, third panel). In the bottom panel the \vs\ was measured within circles of radius $R_{\it e}$. The \vs\ values are now significantly below the isotropic line, particularly for very flattened models, and this shows that in the case of finite spatial coverage the \vs\ should be preferably evaluated along flattened ellipses than along circles.

To test the reliability of the \vse\ diagram in a more realistic case of variable ellipticity, we computed the \vse\ values inside 1\re\ from isotropic models of the five test galaxies computed without changing the ellipticities of the Gaussians in the MGE models of Paper~IV. The ellipticity used is the ellipticity measured on the real galaxies. The result of this test is shown in the second panel of \reffig{fig:isotropic_test}. It shows that the \vse\ values extracted from isotropic models of realistic galaxies, extracted within 1\re, do lie close to the theoretical isotropic line for oblate galaxies (when $\alpha\sim0.2$).

\section{Monte Carlo simulation}
\label{sec:simulation}

\begin{figure}
    \plotone{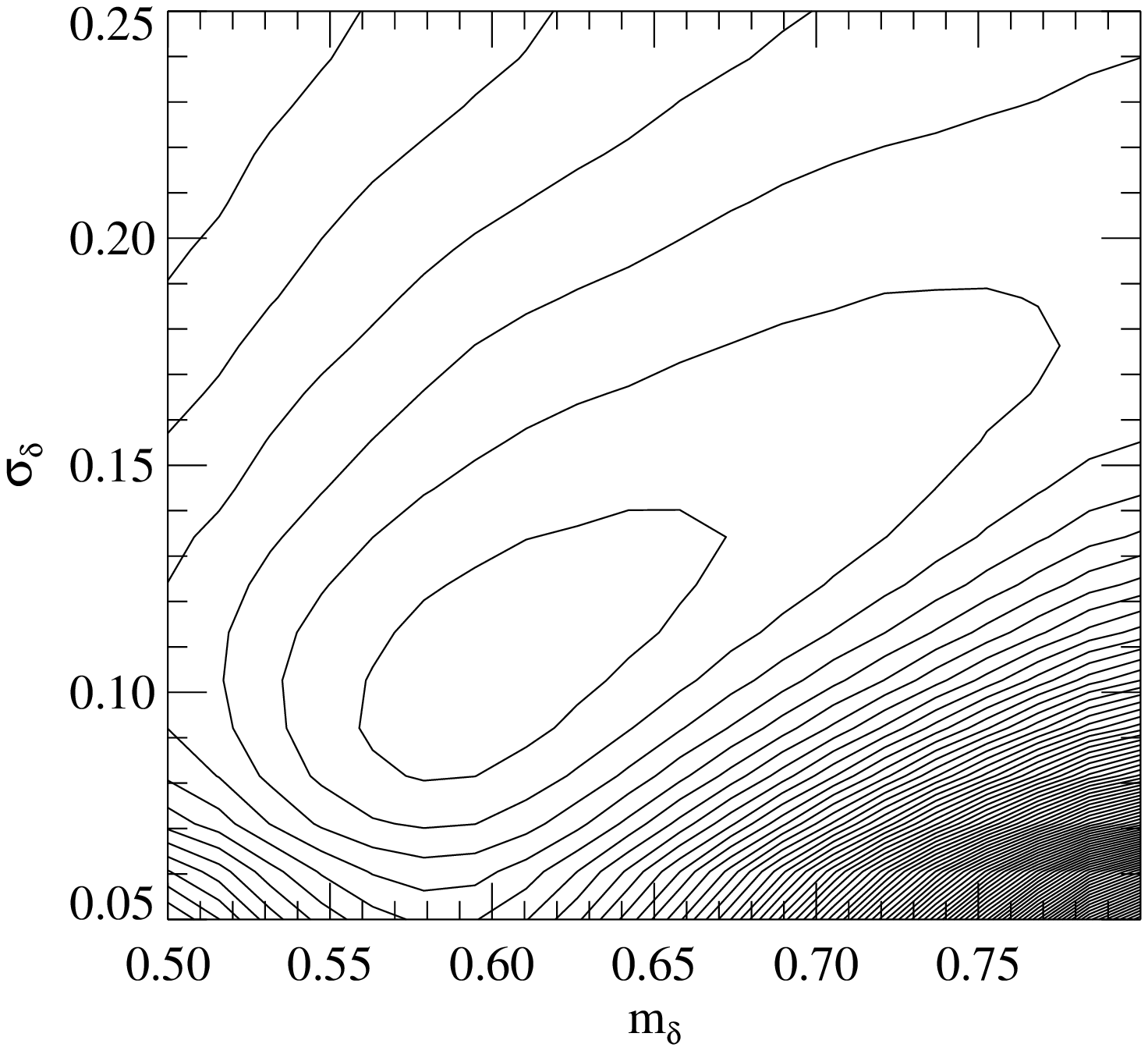}
    \caption{Logarithmically spaced contours of the likelihood $L(m_\delta,\sigma_\delta)$ as a function of the assumed slope and spread of the $\delta-\varepsilon$ correlation. The maximum likelihood values are $m_\delta=0.61$ and $\sigma_\delta=0.11$.}
    \label{fig:anisotropy_best_fit}
\end{figure}

\begin{figure}
    \plotone{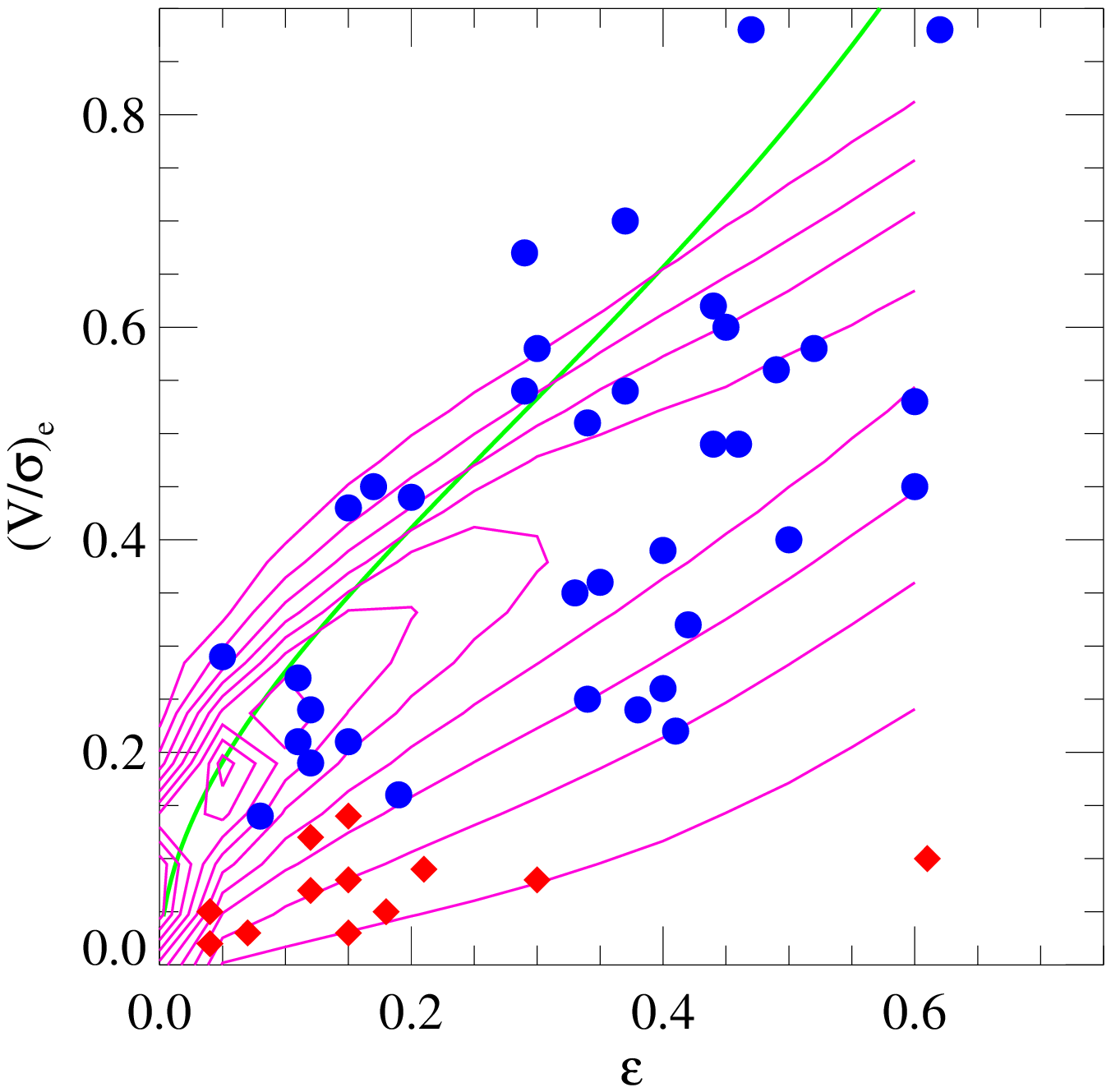}
    \caption{Same as in \reffig{fig:v_over_sigma_lines}, with overlaid with the magenta lines the contour of the probability $P'\left[(\vs)_{\rm obs},\varepsilon; m_\delta,\sigma_\delta\right]$, for the best fitting parameters of \reffig{fig:anisotropy_best_fit}. The contours are meant to describe the distribution of the fast rotating galaxies (blue labels).}
    \label{fig:v_over_sigma_contours}
\end{figure}

\begin{figure}
\centering
\includegraphics[width=0.7\columnwidth]{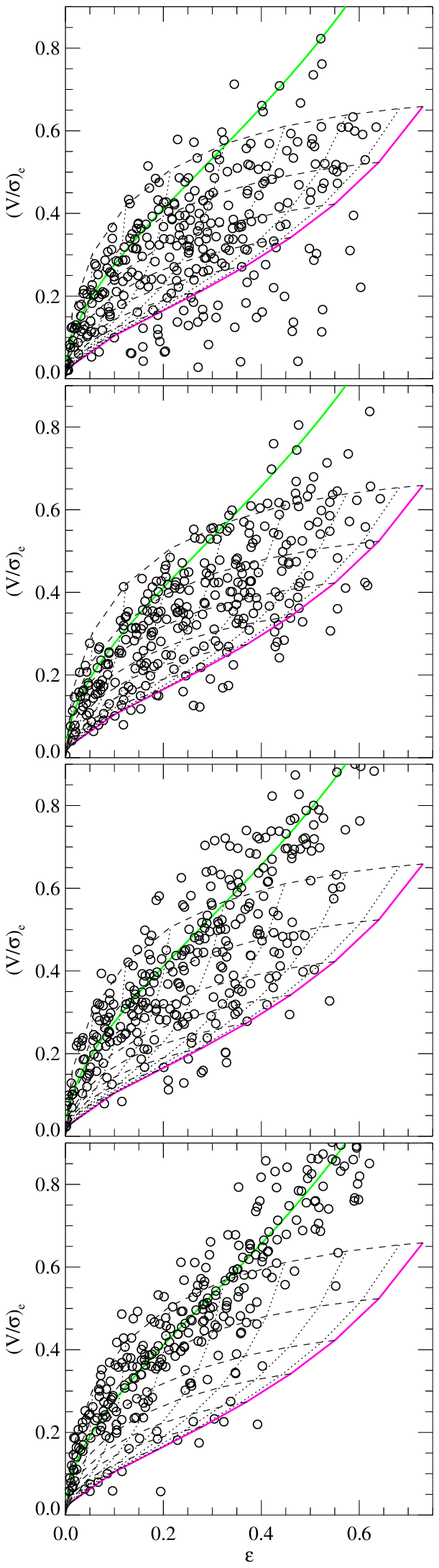}
    \caption{Monte Carlo simulations of the \vse\ diagram. {\em Top panel:} the galaxies follow a relation $\delta=0.6\,\varepsilon_{\rm intr}$ with added Gaussian intrinsic scatter. {\em Second Panel:} same as in top panel, but galaxies with $\delta>0.8\,\varepsilon_{\rm intr}$ are removed. {\em Third Panel:} at every $\varepsilon_{\rm intr}$ galaxies are uniformly distributed in the interval  $\delta=[0,0.8\,\varepsilon_{\rm intr}]$. {\em Bottom Panel:} galaxies are on average isotropic $\delta=0$, with some added Gaussian intrinsic scatter. Only the second panel resembles the observed distribution of the fast-rotators.}
    \label{fig:v_over_sigma_simulation}
\end{figure}

In \refsec{sec:vs_lines} we showed that the location occupied by the fast rotating galaxies in the \sauron\ representative sample is consistent with the one expected according to \refeq{eq:beta_corr}. It is however unlikely that galaxies will follow a correlation with zero scatter. Moreover the previous discussion does not make use of the information contained in the whole distribution of observed galaxies in the \vse\ diagram. As we do not know the inclination for all the galaxies in the sample, the only way to verify if the observed measurements imply a correlation between shape and anisotropy is to resort to Monte Carlo simulations.

We saw in \refsec{sec:shape} that the fast rotators are consistent with being a population of nearly oblate galaxies. In principle, under the oblate hypothesis, it is possible to invert the observed ellipticity distribution to derive the intrinsic shape distribution, under the assumption or random orientations. However the \sauron\ sample was explicitly selected to cover a nearly uniform range of ellipticities (Paper~II), and this means that: (i) one cannot assume random orientations and recover the intrinsic distribution, but also that (ii) the observed distribution on the \vse\ diagram will be weakly dependent on the assumed ellipticity distribution, once the same selection criteria are applied. For this reason, while still assuming the fast rotators are oblate, we will adopt the parametric form for the distribution of intrinsic ellipticities derived by \citet{lam92} for a sample of $\sim5,000$ S0 galaxies:
\begin{equation}
 \phi(\varepsilon_{\rm intr})=\left\{ \begin{array}{ll}
\frac{1}{\sigma_\varepsilon\sqrt{2\pi}}\exp\left[-\frac{(\varepsilon_{\rm intr}-\varepsilon_0)^2}{2\sigma_\varepsilon^2}\right] &  \textrm{if $0\le\varepsilon_{\rm intr}\le0.65$}    \\
 0 & \textrm{otherwise},  \\
 \end{array} \right.
\end{equation}
with $\varepsilon_0=0.41$ and $\sigma_\varepsilon=0.24$, and where we set an upper limit on $\varepsilon_{\rm intr}$ based on the maximum ellipticity observed in our sample.

For any given ellipticity we {\em assume} a simple linear relation with slope $m_\delta$ to determine the anisotropy, as found in \refsec{sec:correlation}, but we allow for an intrinsic (Gaussian) scatter around the relation. The distribution of the anisotropy $\delta$ at any given ellipticity is then assumed to be:
\begin{equation}
\phi(\delta)=\left\{ \begin{array}{ll}
\frac{1}{\sigma_\delta\sqrt{2\pi}}\exp\left[-\frac{(\delta-m_\delta\times\varepsilon_{\rm intr})^2}{2\sigma_\delta^2}\right] &  \textrm{if $0\le\delta\le\frac{\Omega-1}{\Omega}$}    \\
 0 & \textrm{otherwise},  \\
 \end{array} \right.
 \label{eq:delta_mc}
\end{equation}
where the upper limit for $\delta$ comes from the condition $(\vs)^2\ge0$ in equation~(\ref{eq:delta_vs}) and $(m_\delta,\sigma_\delta)$ are two free parameters.

Given $\varepsilon_{\rm intr}$ and $\delta$ we use the inverse of equation~(\ref{eq:delta_vs}) to compute the edge-on \vs\ value. Finally with equations~(\ref{eq:corr_vs},\ref{eq:corr_eps}) we project the $(\vs,\varepsilon_{\rm intr})$ pair of values to get the observed $[(\vs)_{\rm obs},\varepsilon]$ and $(\vs)_{\rm obs}$ values, assuming a uniform distribution on the sphere of the viewing angles (i.~e.\ a uniform probability distribution in $\sin i$). This process, repeated for a large number $\sim10^8$ of random realisations of the of the $\varepsilon_{\rm intr}$ variable, is used to compute an histogram. This constitute a numerical estimate of the probability $P[(\vs)_{\rm obs},\varepsilon; m_\delta,\sigma_\delta]$ of observing a galaxy within a certain range of values on the \vse\ diagram, for any given set of our two model parameter $(m_\delta,\sigma_\delta)$.

One complication now arises because of the \sauron\ selection criteria. We need to consider the fact that the value of $\varepsilon_{\rm obs,j}$ was specifically chosen for each sample galaxy, and does not come from a random selection. For this reason we define a new probability
\begin{eqnarray}
\lefteqn{P'[(\vs)_{\rm obs},\varepsilon; m_\delta,\sigma_\delta] = {}} \nonumber\\
&  &  \frac{P[(\vs)_{\rm obs},\varepsilon; m_\delta,\sigma_\delta]}
{\int_0^\infty P[(\vs)_{\rm obs},\varepsilon; m_\delta,\sigma_\delta]\, \ud (\vs)_{\rm obs}}
\end{eqnarray}
in such a way that
\begin{equation}
\int_0^\infty P'[(\vs)_{\rm obs},\varepsilon; m_\delta,\sigma_\delta]\, \ud (\vs)_{\rm obs} = 1,
\end{equation}
or in other words the marginalised probability of observing a galaxy with any given $\varepsilon$ is constant. What varies is only the probability of observing a certain $(\vs)_{\rm obs}$ value at the chosen $\varepsilon$.

To find the best fitting parameters of our simple model we then define a likelihood function, which gives the probability (neglecting the effect of measurements errors) of observing the given set of $M$ independent measurements $\left[(\vs)_{{\rm obs},m},\varepsilon_{{\rm obs},m}\right]$, for the assumed model parameters $(m_\delta,\sigma_\delta)$:
\begin{equation}
L(m_\delta,\sigma_\delta)=\prod_{m=1}^{M} P'\left[(\vs)_{{\rm obs},m},\varepsilon_{{\rm obs},m}; m_\delta,\sigma_\delta\right].
\label{eq:likelihood}
\end{equation}
We determine the likelihood only for the fast rotators of the \sauron\ sample of 48 galaxies, and excluding the galaxies NGC~2685, NGC~3156, as they lie in regions of low-number statistics and may not follow the general distribution of the other galaxies, and NGC~4473, for the reasons explained in \refsec{sec:vs_lines}. This leads to a sample of $M=36$ galaxies.
The contours of the likelihood function~(\ref{eq:likelihood}) are presented in \reffig{fig:anisotropy_best_fit} and show that the maximum likelihood parameters are $m_\delta=0.6$ and $\sigma_\delta=0.1$. Given the very different approach adopted for their determination, the agreement with \refeq{eq:beta_corr} provides a strong confirmation of the observed trend of anisotropy versus ellipticity. The contours of the probability $P'$ for the set of best fitting parameters are shown in \reffig{fig:v_over_sigma_contours}. Although the contours follow the general distribution of the fast-rotators, they also display a tail with non-negligible probability at low values of \vs, where we do not observe any galaxy. This effect is also visualized in the top panel of \reffig{fig:v_over_sigma_simulation}, where we plot the distribution of values on the \vse\ diagram, for a simulated set of 300 galaxies. A number of galaxies are predicted by this model to fall below the magenta line.

After some experimentation we found that the simplest way to reproduce the observations consists of enforcing the upper limit on the anisotropy $\delta\la0.8\,\varepsilon_{\rm intr}$ in equation~(\ref{eq:delta_mc}). The result of this simulation is shown in the second panel of \reffig{fig:v_over_sigma_simulation} and qualitatively resembles the appearance of the observed \sauron\ \vse\ diagram. The distribution of the observed galaxies in the \vse\ diagram suggests that the above condition constitutes a physical zone of avoidance for the existence of the fast-rotators. This zone of avoidance may be related to the stability of axisymmetric stellar systems with varying degrees of anisotropy and would deserve further investigations.

In \reffig{fig:v_over_sigma_simulation} we present two additional Monte Carlo simulation, under simple assumptions for the galaxy anisotropy. In the third panel we assume, at every $\varepsilon_{\rm intr}$, an uniform distribution for galaxies in the interval $\delta=[0,0.8\,\varepsilon_{\rm intr}]$. This distribution is unable to reproduce the observed distribution of the fast-rotators on the \vse\ diagram. Finally, in the bottom panel we assume that galaxies are on average isotropic $\delta=0$, but we allow for the same Gaussian intrinsic scatter $\sigma_\delta=0.1$ as in the top panel. This distribution is strongly excluded by the observations. A comparison with the \vse\ distribution for a much larger and unbiased sample of galaxies would be required to further confirm these results.

\label{lastpage}

\end{document}